\documentclass[a4paper,11pt]{article}
\pdfoutput=1
\usepackage{jheppub}
\usepackage{float}
\usepackage[T1]{fontenc} 
\usepackage{tikz}

\usepackage{graphicx, color, subcaption, xcolor, float}

\usepackage{bm, amsmath, amsfonts, textcomp, relsize, booktabs, multirow}
\usepackage{gensymb, amsthm, thmtools, tensor,
 amssymb}

\usepackage{fancyhdr}

\allowdisplaybreaks

\title{\boldmath Holographic superconductivity in Einsteinian Cubic Gravity}

\author[a,b]{Jos\'e D. Edelstein,}
\author[c,d]{Nicol\'as Grandi}
\author[a,b]{and Alberto Rivadulla S\'anchez}

\affiliation[a]{Departamento de F\'\i sica de Part\'\i culas, Universidade de Santiago de Compostela,\\
E-15782 Santiago de Compostela, Spain}
\affiliation[b]{Instituto Galego de F\'\i sica de Altas Enerx\'\i as (IGFAE), Universidade de Santiago de Compostela,\\
E-15782 Santiago de Compostela, Spain}
\affiliation[c]{Departamento de F\'\i sica, Universidad Nacional de La Plata,\\
Calle 49 y 115 s/n, CC67, 1900 La Plata, Argentina}
\affiliation[d]{Instituto de F\'isica de La Plata, CONICET\\
Diagonal 113 e/63 y 64, CC67, 1900 La Plata, Argentina}

\emailAdd{jose.edelstein@usc.es}
\emailAdd{grandi@fisica.unlp.edu.ar}
\emailAdd{alberto.rivadulla.sanchez@usc.es}

\abstract{
We study the condensation of a charged scalar field in a $(3+1)$-dimensional asymptotically AdS background in the context of Einsteinian cubic gravity, featuring a holographic superconductor with higher curvature corrections corresponding to a CFT with a non-vanishing value of the stress tensor three-point function $t_4$. As it was previously noticed for higher dimensional Gauss-Bonnet theory, we observe that the critical temperature of the superconducting phase transition is lowered as the higher curvature coupling grows.}

\begin{document}
\maketitle

\section{Introduction}
\label{sec:introduction}

Holography has become a standard tool to study the physics of quantum field theories at strong coupling. There is a rich class of holographic models which describe low dimensional strongly coupled quantum field theories at finite density. Since such setup can be taken as a proxy for condensed matter systems, the resulting branch of research is known as AdS/CMT for Condensed Matter Theory \cite{AdSCMTbook}. A particularly interesting AdS/CMT construction is the one dubbed ``holographic superconductor'' \cite{HHH-p1, HHH-p2}. It describes a superconducting phase at low temperature and magnetic field, in which a $U(1)$ symmetry is spontaneously broken \cite{Tinkham}. Several universal features of the laboratory systems known as High T$_c$ superconductors are mimicked by their holographic analogs \cite{Hartnoll-book}.

A basic ingredient of the holographic duality is that, when the dynamics of the bulk is classical, the boundary field theory has a large number of degrees of freedom at any point in space \cite{Maldacena}. This is not realistic enough if one wants to push further the analogy between holographic and High T$_c$ superconductors. Such ``large $N$ limit'' can be improved by calculating loop diagrams in the bulk, which introduce $1/N$ corrections into the boundary results. A subset of such diagrams can be resummed in the form of higher curvature corrections to the bulk gravitational action \cite{GrossWitten}. In consequence, classical holography in the resulting modified gravity backgrounds takes into account some of the $1/N$ corrections ---as well as $1/\sqrt{\lambda}$, where $\lambda$ is the 't Hooft coupling--- to the dual field theory.

The precise form of the higher curvature corrections depends on the field content and the regularization scheme. In a top-down scenario, this depends in turn on the D-brane construction, an intricate avenue that is customarily circumvented by choosing a bottom-up approach, where generic physical requirements can be imposed in order to build a ``healthy'' higher order gravitational theory. In particular, the absence of ghosts in maximally symmetric backgrounds and the existence of black-hole solutions characterized by a single radial function result, at third order in curvature, in a correction to the Einstein-Hilbert Lagrangian known as ``Einsteinian Cubic Gravity'' (ECG) \cite{ECG},
%
%
and generically in a set of models called  ``Generalized Quasitopological Gravities'' (GQTGs) \cite{GQTG}. If we further require a cosmological scenario ruled by second order field equations, we are led to a unique cubic 
%
correction
accomplishing all the aforementioned conditions \cite{Arciniega:2018fxj}, 
\begin{eqnarray}
S_{\sf Einsteinian}^{(3)} &=&\frac{\tilde\beta L^4}{108\kappa^2}  \!\int \!d^4x\left( R_{ab}{}^{cd} R_{cd}{}^{ef} R_{ef}{}^{ab} + 12 R_{a}{}^{c}{}_{b}{}^{d} R_{c}{}^{e}{}_{d}{}^{f} R_{e}{}^{a}{}_{f}{}^{b} - 8 \tensor{R}{_a_b_c_d} \tensor{R}{^a^b^c_e} \tensor{R}{^d^e}
\right. \nonumber\\ [0.9em] & & \left. \qquad\,
 + \,2 R_{abcd} R^{abcd} R  + 4 R_{abcd} R^{ac} R^{bd} + 8 R_{a}{}^{b} R_{b}{}^{c} R_{c}{}^{a} - 4 \tensor{R}{_a^b} \tensor{R}{_b^a} R \right),
\label{LagrangianR3}
\end{eqnarray}
%
where $\kappa^2$ is the gravitational coupling, $L$ is the AdS length, and $\tilde\beta$ is a parameter controlling the coupling of the Einsteinian cubic terms.
Interestingly enough, this can also be generalized to all orders \cite{Arciniega:2018tnn}. In the present holographic context the above requirement on the cosmological setup can be mapped, via a double Wick rotation interchanging cosmological time with the holographic coordinate, into the statement that Lorentz invariant boundary scenarios are extended to the bulk by solving second order field equations.

In $(2+1)$-dimensional systems at finite temperature, Coleman-Mermin-Wagner theorem rules out the spontaneous breaking of a continuous symmetry, due to the presence of long-wavelength fluctuations \cite{MerminWagner, Coleman}. In holographic systems, these fluctuations are suppressed in the large $N$ limit \cite{AGMN-SG}, corresponding to Einstein gravity, thus allowing for the condensation of a scalar field during a phase transition. 
Being an infrared effect, Coleman-Mermin-Wagner result would not be restored by merely adding $1/N$ corrections to the bulk classical action, and would require the calculation of Witten loops in the bulk. However, it has been observed that higher curvature couplings make the scalar condensation more difficult, delaying the superconducting phase transition \cite{Gregory, Pan, Kuang, Barclay}. Since these works deal with higher dimensional systems, Einsteinian cubic gravity opens the posibility to investigate the presence of such effect on a $(2+1)$-dimensional boundary theory. 

It is the aim of the present work to study the finite $N$ and finite 't Hooft coupling corrections to the holographic superconductor using the aforementioned Einsteinian cubic bulk action for the gravitational degrees of freedom. This action describes a dual CFT with a non-vanishing value for the stress tensor three-point function $t_4$ \cite{Holographic studies of ECG}. We concentrate in the $s$-wave superconductor, whose condensate is dual to a charged scalar field in the bulk.
 
The paper is organized as follows. In section \ref{sec Model} we present our holographic model, including the action, the holographic Ansatz, and the resulting equations of motion. In section \ref{sec Normal} we solve for the normal phase, which corresponds to the Einsteinian cubic gravity charged black hole. In section \ref{sec Superconductor} we explore the superconducting phase, first in the proble limit studying the condensation of the scalar operator and the corresponding conductivity, and then in the fully backreacting setup. Finally in section \ref{sec Discussion} we present our conclusions. Some technical material on the numerical method and the near horizon geometry is provided in  Appendices \ref{appendix Numerical relaxation method for differential equations} and \ref{appendix AdS2}.

\newpage
\section{The model}
\label{sec Model}

We consider a holographic model for a $s$-wave superconductor with a $(2+1)$-dimensional boundary. The standard procedure is to supplement the gravitational sector with a $U(1)$ gauge field that extends to the bulk the boundary global $U(1)$ symmetry, and a charged scalar to represent the condensate that would eventually break it in the superconducting phase. This leads to the following $(3+1)$-dimensional bulk action
\begin{equation}
S = S_{\sf Gravity} + S_{\sf Maxwell} + S_{\sf Scalar}\,,
\label{eq action model}
\end{equation}
where the terms for the electromagnetic and charged scalar fields have the standard form
\begin{equation}
S_{\sf Maxwell} + S_{\sf Scalar}=
-\int d^4x \, \sqrt{-g} \left( \frac{1}{4}F_{ab}F^{ab} +|\partial\psi - i { q} A\psi|^2 + m^2 |\psi|^2 \right),
\end{equation}
here $m$ and $q$ are the scalar field's mass and charge respectively. 
For the gravity dynamics on the other hand, we include the Einsteinian cubic corrections \eqref{LagrangianR3} to write it in the form
\begin{equation}
S_{\sf Gravity} = \frac{1}{2\kappa^2} 
\int d^4x \, \sqrt{-g} \left( R +\frac{6}{L^2}  
\right) +S^{(3)}_{\sf Einsteinian}\,.
\label{eq 2 Gravity action third order general}
\end{equation} 
%
%
%

The action (\ref{eq action model}) has a well defined probe limit, in which the scalar and electromagnetic fields do not curve the background geometry. This happens whenever the fields $\psi$ and $A$ and their derivatives are small, since the corresponding energy momentum tensor is quadratic in those fields, while their equations of motion contain linear terms. To keep some interaction between the scalar and electromagnetic sectors in this limit, we need to take $q\to\infty$ while keeping $q A$ and $q\psi$ finite. In the forthcoming sections we study the system in both the probe limit and the backreacting regime. 

Since we are interested in a spatially infinite boundary system at equilibrium, we look for solutions wich are static and have AdS asymptotics, with a Euclidean 2-dimensional spatial symmetry on the boundary. This is realized by the planar Ansatz
\begin{align} \label{eq 2 Ansatz planar black hole}
&ds^2 = - n^2(r)F(r) dt^2 + \frac{dr^2}{F(r)} + \frac{r^2}{L^2}(dx^2 + dy^2)\,,
\\
&\psi= \psi(r)\,,
\qquad\qquad\quad
A = \phi(r) \,dt\,.
\label{eq 2 ansatz planar}
\end{align}
The equations of motion for the coupled fields can be obtained by using the reduced action approach, this is, evaluating the action with the Ansatz and taking the functional derivative with respect to each of the functions in the fields. After setting $2 \kappa^2 = 1$ 
for simplicity, they read  
%
\vspace{.5cm}
\begin{subequations}
\begin{align}
&
\psi'' + \left(\frac{2}{r}+\frac{n'}{n} + \frac{F'}{F}\right)\psi' + \frac{1}{ F}\left( \frac{q^2 \phi ^2}{n^2 F}-m^2 \right) \psi = 0\,,
\label{eq:scalar}
\\\nonumber & 
\\ & 
\phi'' + \left( \frac{2}{r} - \frac{n'}{n}\right) \phi' - \frac{2 q^2 \psi^2}{F} \phi = 0\,, \label{eq:gauge}
\\\nonumber & 
\\\nonumber & 
F n r^3 
\Big[
	2r n^2 
	\left(
		6 \frac{r}{L^2} - m^2r \psi ^2-2F'
	\right)
	-2F n^2
	\left(
		r^2 \psi '^2	+ 2
	\right)	
	-r^2  \phi '^2
\Big]
-2 q^2 r^5 n \psi ^2 \phi ^2
\\\nonumber & 
+ 
\frac 2{27} \,\tilde\beta L^4
\bigg\lbrace \phantom{2^{2^2}}\!\!\!\!\!\!\!
24 F^2 n^3 F' (rF' - F ) - 24 F^4 n' \big(2(r^2 n'^2-n^2)+r n n' \big)
\\\nonumber&\hspace{1.5cm} 
+6rFF'n'\Big[ 8 F n^2 (rF'-2F)- r^2 n F'(5 F n'+2 n F') +4rF^2n'(4  n  + r  n' ) \Big] 
\\\nonumber&\hspace{1.5cm} 
+6rnF^2F'' \Big[ 4 F\big( - r n^2 F' +( n^2 + r nn' - r^2  n'^2 ) \big)+ r^2  n^2 F'' \Big] 
\\\nonumber&\hspace{1.5cm} 
+12rF^2nn'' \Big[ 2F^2(3 r n' - n)+rF' \big( n (2 F - r   F') -3 r  n'F \big) + r^2 n F F'' \Big] 
\\ &\hspace{1.5cm} 
+3r^2n^2 F^3F^{(3)} \big( 4 (r n'-  n) + 2 r F^2 n F' \big)
\bigg\rbrace=0\,,
\label{eq:einstenian1}
\\\nonumber &
\\\nonumber & 
nr^3 \Big( 2 F^2 n n'-r F^2 n^2 \psi '^2-q^2 r \psi ^2 \phi ^2 \Big)
\\\nonumber & 
+\frac 2{27} \,\tilde\beta L^4 \bigg\lbrace
3n' \Big[
4 n^2F^2 F' (r  F' -2 F    ) +F^2  n' \big( 4 F^2  (n -2 r  n')
+3 r^2 F' ( 2 F n'  -n F') \big) \Big]
\\\nonumber & \hspace{1.5cm} 
+ 3 nF^2 n'F'' \Big[ nr(8 F-5 rF')-6  r^2 F n' \Big]
\\\nonumber & \hspace{1.5cm} 
+3F^2n'' \Big[ 2n' F \big( 2nr (F -3 r  F')+ r^2 F n' \big)  
+n^2  \big( 2F ( 4 r   F' - F ) -3  r^2 F'^2 \big) 
 \Big]
 \\	& \hspace{1.5cm}
- 3r^2nF^3 n'' \big( 2F  n''  
+ n F'' \big) 
- 3rnF^3 n^{(3)} \big( 2 F(r  n' - n)+ r n F' \big)
\bigg\rbrace = 0\,.\label{eq:einstenian2}
\end{align}
\label{eq ECG EOMs original fields}%
\end{subequations}

Asymptotically AdS solutions have the large $r$ metric expansion
\begin{align}
F(r) & =
\frac{r^2}{L^2}
f_\infty
+{\cal O}\left(r^{-1}\right)\,,
\label{exp1F}
\\
n(r)& = n_\infty +
 {\cal O}\left( r^{-2} \right)\,,
\label{exp1n}
\end{align}
%
where $f_\infty$ and $n_\infty$ 
are constants. We re-scale the time variable $t$ such that $n_\infty^2f_\infty=1$, setting it to measure the boundary time. On the other hand, finite temperature configurations have a horizon at finite $r=r_h$ where $F(r)$ vanishes. There, the metric functions can be expanded according to
\begin{align}
F(r) & = \frac{4\pi T}{n_h} (r-r_h) 
+{\cal O}(r-r_h)^2\,,
\label{exp2F}
\\
n(r)& = n_h+{\cal O}(r-r_h)\,,
\label{exp2n}
\end{align}
where $n_h$ is a constant that sets the time units at the horizon, and $T$ is the black hole temperature. The above expressions, supplemented by suitable expansions of $\psi(r)$ and $\phi(r)$, can be plugged into the equations of motion to relate the constants and obtain the subleading terms. We will do that independently on each phase in the forthcoming sections. 

\newpage
\section{The normal phase}
\label{sec Normal}
The normal phase is obtained using the Ansatz (\ref{eq 2 ansatz planar}) with $\psi$ and $\phi$ vanishing identically. 
Then equations \eqref{eq:einstenian1}-\eqref{eq:einstenian2} simplify and can be partially integrated, obtaining
\begin{subequations}
\begin{align} 
\frac{r^3}{L^2} - F \,r +  \frac {L^4 }{27} \,\tilde\beta \left(3 F F' F''  -F'^3 + \frac{6 }{r^2} F^2 \left(F' - r F'' \right) \right) &= M\,,
	\label{eq:integralF}
\\
	n(r)&=n\,,
	\label{eq:integraln}
\end{align}
\end{subequations}
where we introduced two integration constants: $M$ which is proportional to the mass of the black hole \cite{Holographic studies of ECG} and $n$ which sets the time units.  

Plugging the asymptotic expansions \eqref{exp1F}-\eqref{exp1n} back into the equations of motion, we obtain  $n=n_\infty =1/\sqrt{f_\infty}$, together with the relation
\begin{align}
& f_\infty\left(1-
\frac{4}{27}\tilde\beta
f_\infty^2\right)=1\,.
\label{eq 2 Relations m_eff f_infty}
\end{align}
To protect the signature of the metric we need solutions 
$f_\infty > 0$, and they exist whenever $\tilde\beta \leq 1$. 
On the other hand, higher terms in the expansion are well behaved \cite{Holographic studies of ECG} whenever $\beta \geq 0$. 
Then we get the constraint $0 \leq\tilde \beta \leq 1$. Moreover, in this range we have two branches of solutions of \eqref{eq 2 Relations m_eff f_infty} with $f_\infty$ positive. In order to have a positive effective Newton constant for the perturbations of the black-hole, we select the branch that flows into Einstein AdS black hole when $\tilde\beta$ goes to zero \cite{Bestiary}, or in other words the one that satisfies $\lim_{\tilde\beta \rightarrow 0} f_\infty = 1$. Then we have
\begin{equation}
	f_\infty(\beta) = \frac{3}{\sqrt{\tilde{\beta}}} \sin \left[ \frac{1}{3} \arcsin \left( \sqrt{\tilde\beta} \right) \right].
	\label{eq 2 f_infinity analytic expression}	
\end{equation}
In consequence, the resulting solution is completely determined in terms of the integration constant $m$ and the cubic coupling $\tilde\beta$.

At the horizon, we have expansion (\ref{exp2F})-(\ref{exp2n}), which when plugged in the equation of motion for $F(r)$ and imposing that the terms multiplied by each power of $r-r_h$ vanish independently, implies
\begin{equation}
	T = \frac{3 r_h}{4 \pi L^2 \sqrt{f_\infty}}, \qquad r_h = \left( \frac{M L^2}{1 - \tilde \beta} \right)^{1/3},
	\label{eq 2 Temperature horizon radius with m}
\end{equation}
determining the temperature and horizon position. 

There is a value of the cubic coupling, $\tilde\beta = 1$, that makes the effective Newton's constant divergent. This limit is interesting from the computational point of view, since the black hole solution becomes analytical,
\begin{equation}
	F(r) = \frac{3}{2 L^2} \left( r^2 - r_h^2 \right).  
	\label{eq 2 Critical solution f(r)}
\end{equation}
However, this solution corresponds to $M = 0$, which means that the mass of the black hole vanishes. Also, the horizon radius $r_h$ and the temperature are not determined by the equations of motion.

For generic values of $\tilde\beta$, we need to compute the form of the function $F(r)$ outside the horizon.  For $\tilde \beta < 1$ the equations of motion cannot be solved analytically, so we need to resort to numerical methods. 
In \cite{4 dim BHs in ECG,Holographic studies of ECG}, the authors implement a numerical shooting method to find the solution. While this method is able to produce solutions with arbitrary numerical accuracy, it becomes much more complex when dealing with backreaction of the matter field, as we plan to do in the forthcoming sections. This happens because the equation of motion for $F(r)$ is not second order in general and $n(r) \neq \text{constant}$. Therefore, we employ a numerical relaxation method, described in Appendix \ref{appendix Numerical relaxation method for differential equations}, its main advantage being that it becomes easier to impose boundary conditions simulaneously at the two ends of the integration interval.

\begin{figure}[h!]
\vspace{.5cm}
	\centering
	\includegraphics[width=0.65\linewidth]{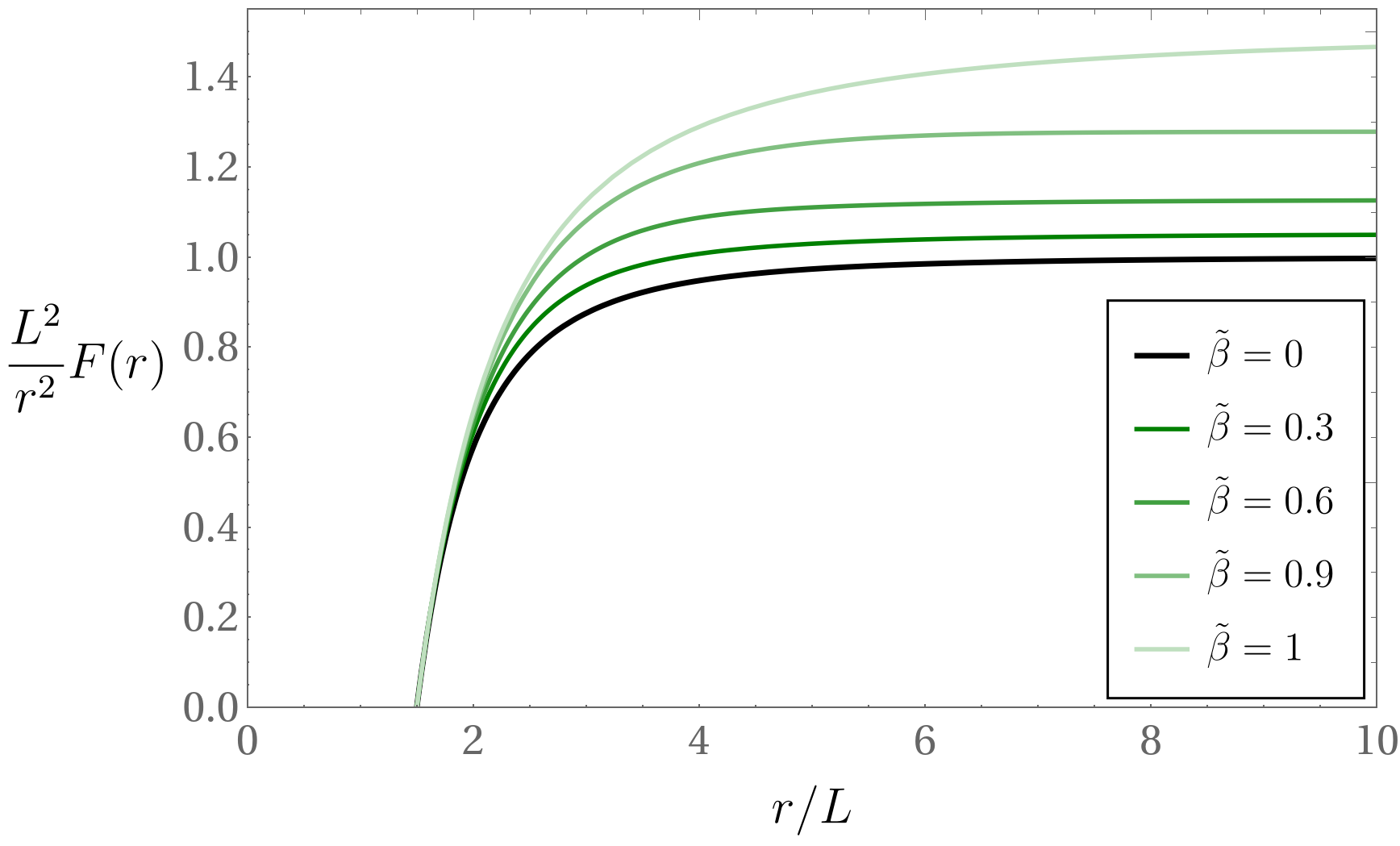}
	\caption{Numerical solution of $F(r)$ for several values of $\tilde\beta$ and $r_h = 1.5 L$, using the relaxation method with 40 grid points.}
	\label{fig 2 Numerical solution black hole several beta}
\end{figure}

The numerical solutions found for several values of $\tilde \beta$ are shown in figure \ref{fig 2 Numerical solution black hole several beta}, where we compare them to the result from general relativity and the critical solution (\ref{eq 2 Critical solution f(r)}). The numerical accuracy of the relaxation method was checked by plugging the numerical solution into the equation of motion, and it was found to yield significantly lower errors than the shooting method of \cite{4 dim BHs in ECG,Holographic studies of ECG}. The numerical solutions obtained here also match the exact forms in the cases $\tilde\beta = 0$ and $\tilde\beta \rightarrow 1$, respectively.


\section{The superconducting phase}
\label{sec Superconductor}
At low enough temperature, expressions \eqref{exp2F}-\eqref{exp2n} dictate that the near horizon metric is that of an AdS$_2$ geometry. As in the standard holographic superconductor, choices of the scalar mass that are stable in the asymptotic AdS$_4$ regime, violate instead the two dimensional Breitenlohner-Freedman bound in the near horizon region.  This leads to an instability and into the development of a charged scalar hair, representing the superconducting phase. This is reviewed in the Einsteinian cubic context in Appendix \ref{appendix AdS2}. 

\subsection{Probe limit}
\label{sec Probe}
If we assume that the scalar field $\psi$ and the electric potential $\phi$, as well as their derivatives, are small enough, then they can be discarded in the gravity equations \eqref{eq:einstenian1}-\eqref{eq:einstenian2} where they appear squared. The gravity sector is then effectively decoupled, leading to the first integrals \eqref{eq:integralF}-\eqref{eq:integraln} that we discussed in the previous section. These provide the gravitational background, which describes a neutral Einsteinian black hole, on which the equations for the scalar \eqref{eq:scalar} and gauge potential \eqref{eq:gauge} must be solved. Notice that, in order to keep a non-trivial gauge coupling for the scalar field in this non-backreacting setup, we need to take the limit $q\to\infty$ while keeping $q\psi , \, q\phi\approx$ constant.

At the horizon, we impose regular boundary conditions, in order for $A_\mu A^\mu$ to not diverge. As it can be checked in the equations of motion \eqref{eq:scalar}-\eqref{eq:gauge}, this implies 
\begin{align}
&\phi = 
{\cal O}(r-r_h)\,,
\label{bcphi}
\\
&\psi = \psi_h+{\cal O}(r-r_h)\,,
\label{bcpsi}
\end{align}
where $\psi_h$ is a constant. 
At large $r$ on the other hand we obtain from equations \eqref{eq:scalar}-\eqref{eq:gauge} the asymptotic expansions
\begin{align}
&\phi=\mu+\frac{\rho}{r}+{\cal O}\left(r^{-2}\right)\,,
\label{eq:asympPhi}
\\
&\psi = \psi_+ \left(1+{\cal O}\left(r^{-1}\right)\right)r^{-\Delta_+} + \psi_- \left( 1+{\cal O}\left(r^{-1}\right)\right)r^{-\Delta_-}\,,
\label{eq:asympPsi}
\end{align}
where $\mu,\rho$ and $A_\pm$ are constants, and the powers $\Delta_\pm$ are given by
\begin{equation}
\Delta_\pm=\frac{3}{2}\pm\sqrt{\frac{9}{4}+\frac{L^2m^2}{f_\infty}}\,.
\label{eq:leadingsubleading}
\end{equation}

In order for the solution to be stable near the boundary, the discriminant inside the square root on \eqref{eq:leadingsubleading} needs to be positive, resulting in the well known Breitenlohner-Freedman bound on the scalar mass. Notice that, since according to \eqref{eq 2 f_infinity analytic expression} we have $1 \leq f_\infty \leq 3/2$ the bound is lowered by the cubic curvature terms, this is, the field can have a larger tachyonic mass while still being stable. 

In what follows, and in order to have the convenient values $\Delta_+=2$ and $\Delta_-=1$ to make contact with the standard literature, we choose the mass of the scalar field as $m^2=-2f_\infty/L^2$. This implies that both terms in the expansion \eqref{eq:asympPsi} for the scalar field are normalizable, and one of its coefficients $\psi_+$ (or $\psi_-$) are proportional to the expectation value of a dual boundary operator $\langle{\cal O}_+\rangle$ (respectively $\langle{\cal O}_-\rangle$) according to $\langle \mathcal{O}_\pm \rangle = \sqrt{2} \psi_\pm$. The remaining coefficient $\psi_-$ (or $\psi_+$) can then be identified with the corresponding source ${\cal J}_-$ (respectively ${\cal J}_+$). Since we are interested in spontaneous symmetry breaking, we will set such source to zero.

\begin{figure}[h!]
	\centering
	\begin{subfigure}{0.49\linewidth}
		\includegraphics[scale=0.44]{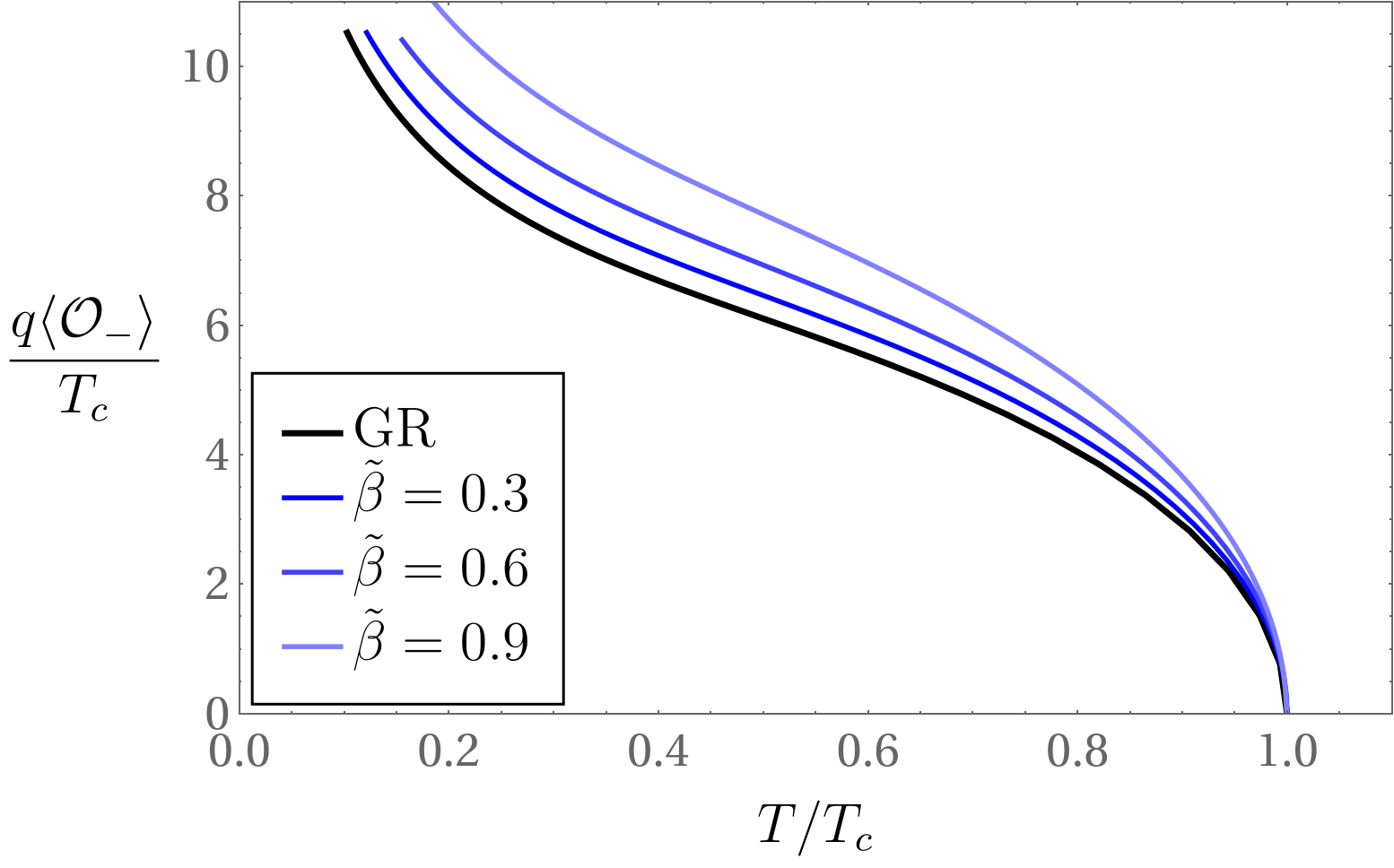}
		\caption*{$\langle \mathcal{O}_-\rangle$ condensate.}
	\end{subfigure}
	\centering
	\begin{subfigure}{0.49\linewidth}
		\includegraphics[scale=0.44]{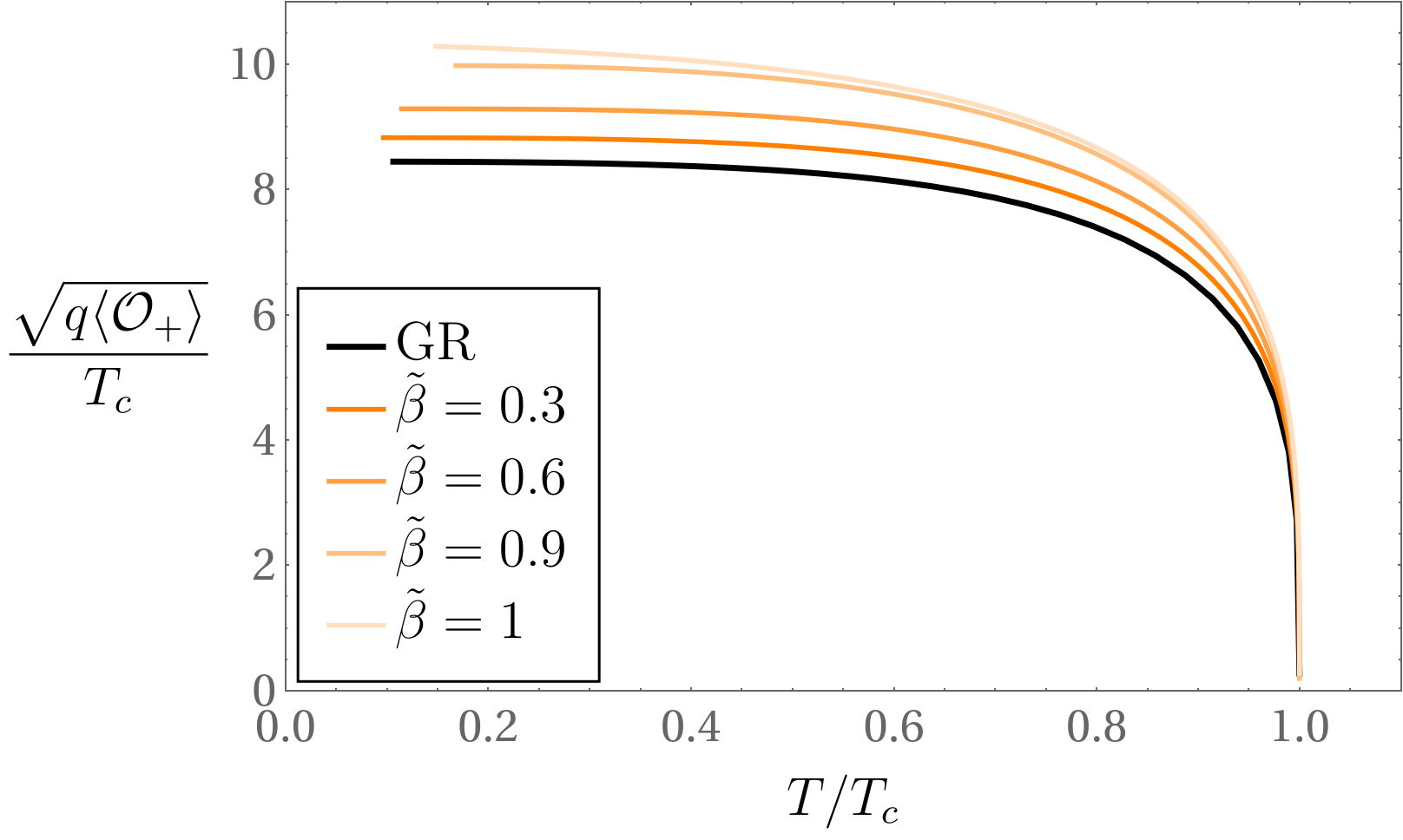}
		\caption*{$\langle \mathcal{O}_+\rangle$ condensate.}
	\end{subfigure}
	\caption{Condensation of the dimension 1 and 2 operators with respect to the temperature in the probe limit ($q \rightarrow \infty$), for different values of $\tilde\beta$.}
	\label{fig ECG Condensation of the operators with respect to the temperature in the probe limit. Different beta}
\end{figure}

In Fig.\ \ref{fig ECG Condensation of the operators with respect to the temperature in the probe limit. Different beta} we see how the condensate value $\langle{\cal O}_\pm\rangle$ in each of the possible quantizations changes as a function of the temperature. The general result is that increasing values of the cubic coupling $\tilde \beta$ leads to larger condensates at low temperatures. This is analogous to what was previously reported in other higher curvature theories in higher dimensions \cite{Gregory, Kuang}. As in the standard GR case, there is a divergence on the $\langle{\cal O}_-\rangle$ condensate as $T$ goes to zero, which spoils the decoupling limit and disappears when backreaction is considered.

\begin{figure}[h]
	\centering
	\includegraphics[width=0.55\linewidth]{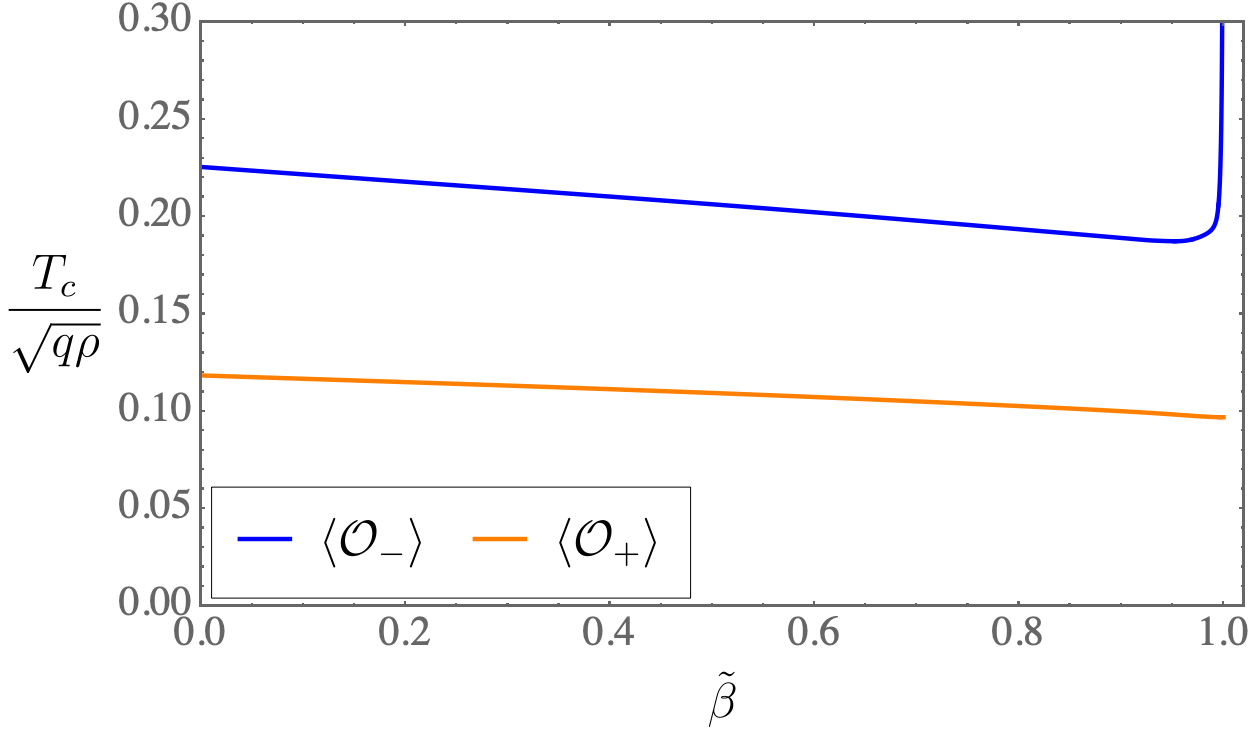}
	\caption{Variation of the critical temperature of the two operators in the probe limit with respect to $\tilde \beta$. Notice that as $\tilde \beta$ is increased, the critical temperature gets lowered. The divergence on the blue curve at $\tilde \beta\to1$ is not due to the critical temperature, but to the vanishing of $\rho$ in the $\tilde \beta=1$ case.}
	\label{fig ECG critical temperature with respect to beta}
\end{figure}

In Fig.\ \ref{fig ECG critical temperature with respect to beta} we see that the critical temperature for the superconducting phase transition decreases monotonically as a function of the cubic coupling, until values of $\tilde\beta$ close to its upper bound, $\tilde{\beta} = 1$, are approached. There, there is a qualitative difference depending on the chosen quatization, which can be explained as follows. In the limit $\tilde{\beta} = 1$ the function $F(r)$ is given exactly by \eqref{eq 2 Critical solution f(r)}, where the radius of the horizon can take any positive value. On the other hand, the asymptotic form of $\phi(r)$ and $\psi(r)$ that defines the normal and superconducting phases is still given by \eqref{eq:asympPhi} and \eqref{eq:asympPsi}. However, if we look at the condensate $\langle\mathcal{O}_-\rangle$, this is, we set $\psi = \psi_- / r$ and $\phi = \mu + \rho / r$, the equations of motion \eqref{eq:scalar} and \eqref{eq:gauge} imply $\psi_- = \rho = 0$, which means that both fields are equal to zero in this limit. Thus, it makes sense that the critical temperature diverges as seen in Fig. \ref{fig ECG critical temperature with respect to beta}, since indeed the energy scale with respect to which it is measured, $\sqrt{\rho}$, vanishes. This does not happen with the other condensate $\langle\mathcal{O}_+\rangle$, in which case $\psi_+$, $\mu$ and $\rho$ are not determined analytically by the equations of motion and the entire numerical computation needs to be carried out.

\subsubsection*{Conductivity}

The electric conductivity of the dual theory can be evaluated using standard holographic methods. To that end, we turn on a perturbation on the spatial component of the gauge field $e^{-i\omega t}\delta A_x(r)$, which results in the linear homogeneous equation 
\begin{equation}
\delta A_x'' + \left(\frac{F'}{F} +\frac{n'}{n} \right) \delta A_x' + \frac 1F\left(\frac{\omega^2}{n^2 F} - {2 q^2 \psi^2}\right) \delta A_x = 0\,.
\label{eq ECG EOM Ax original}
\end{equation}
This would excite a perturbation in the $tx$ component of the metric $e^{-i\omega t}g_{tx}$, but in the probe limit we can safely turn it off.

The behavior of $\delta A_x(r)$ at the horizon reads
\begin{equation}
\delta A_x=\delta A_{\sf out}(r-r_h)^{i\frac \omega{4\pi T}}+\delta A_{\sf in}(r-r_h)^{-i\frac \omega{4\pi T}}\,,
\end{equation}
while on the other hand close to the AdS boundary we get
\begin{equation}
\delta A_x(r) = \delta A_x^{(0)} + \frac{\delta A_x^{(1)}}{r}+{\cal O}(r^{-2})\,.
\label{eq ECG behavior Ax boundary}
\end{equation}
These two asymptotic modes become non-independent when we impose causal boundary conditions at the horizon by setting $\delta A_{\sf out}=0$. With this, we can calculate the conductiviy by identifying the leading term with the perturbation on the electric potential $\delta E_x{\sqrt{f_\infty}}/i\omega$ and the subleading one with its linear response on the electric current $\delta J_x$, according to the Kubo formula
\begin{equation}
\sigma = \frac{\sqrt{f_\infty}}{i \omega} \frac{\delta A_x^{(1)}}{\delta A_x^{(0)}}\,.
\label{eq ECG conductivity formula}
\end{equation}

\begin{figure}[h!]
\vspace{-0.2cm}
	\centering
	\begin{subfigure}{0.44\linewidth}
		\includegraphics[width=\linewidth]{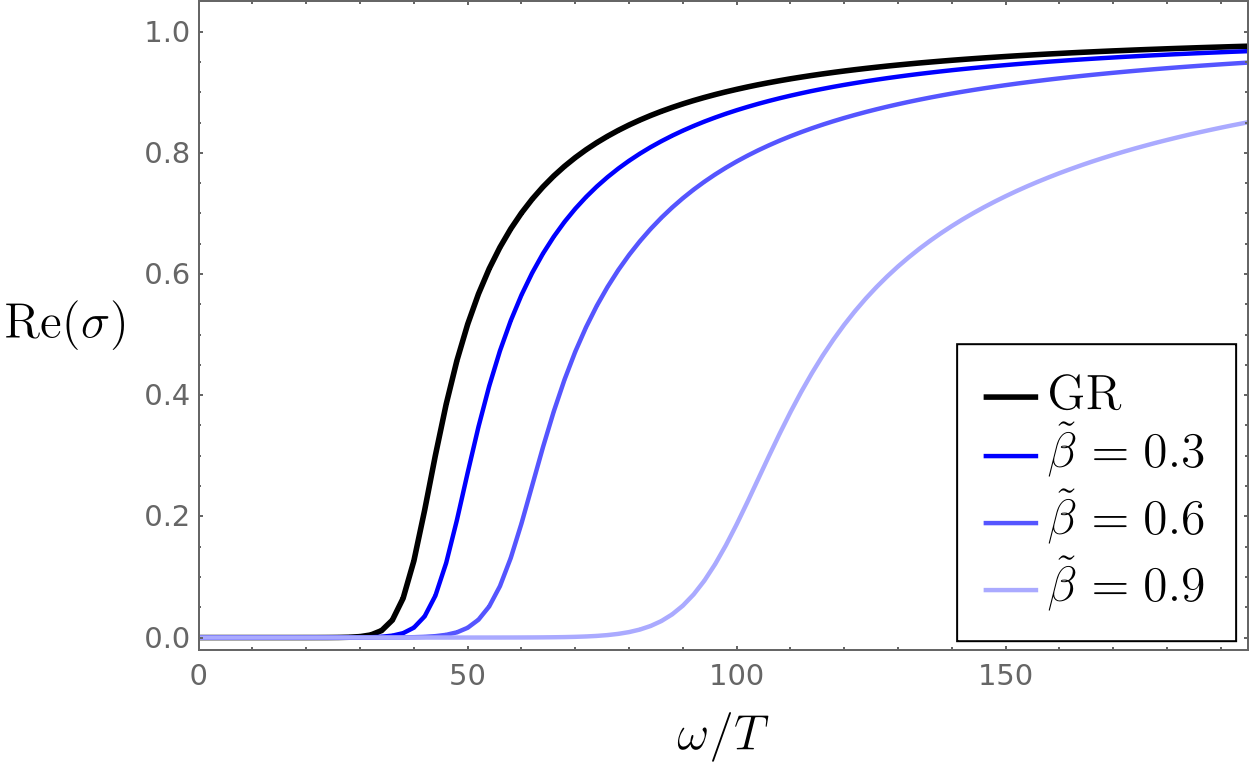}
	\end{subfigure}
	\centering
	\begin{subfigure}{0.44\linewidth}
		\includegraphics[width=\linewidth]{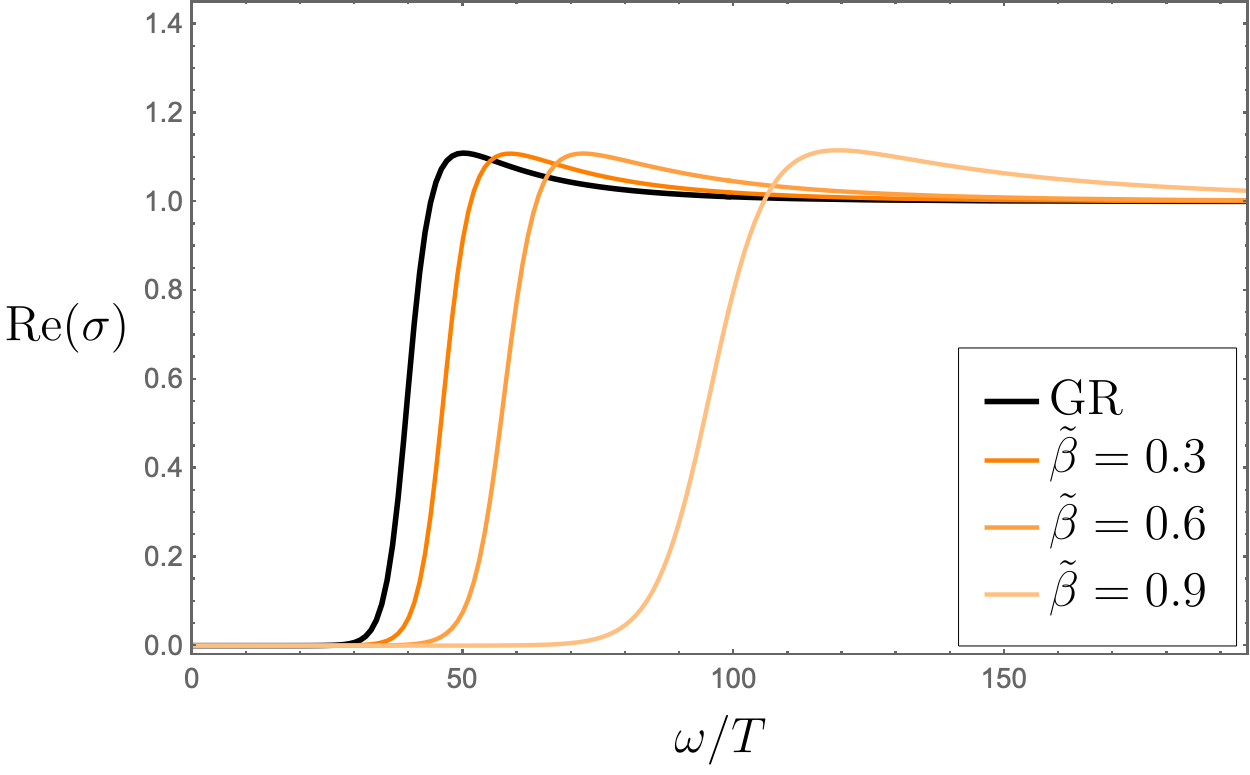}
	\end{subfigure}
	\centering
	\begin{subfigure}{0.44\linewidth}
		\includegraphics[width=\linewidth]{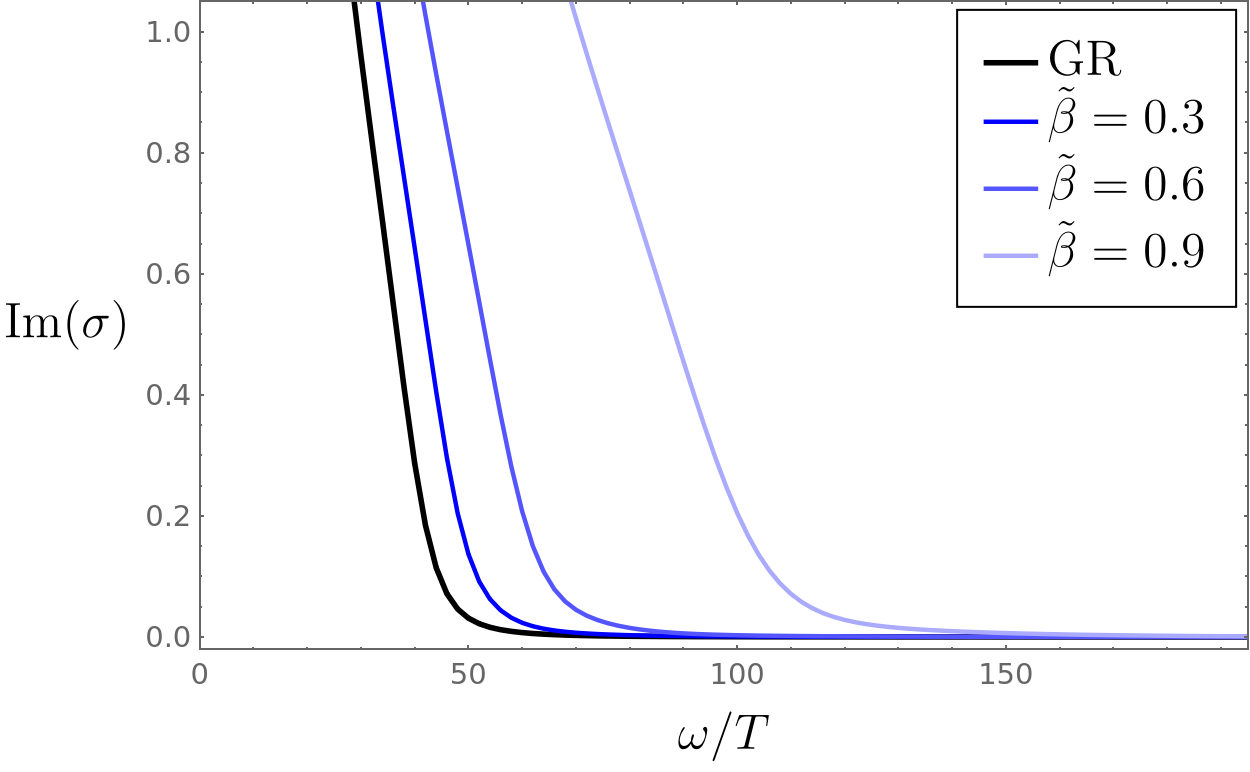}
		\caption*{$\langle \mathcal{O}_-\rangle$ condensate.}
	\end{subfigure}
	\vspace{.7cm}
	\centering
	\begin{subfigure}{0.44\linewidth}
		\includegraphics[width=\linewidth]{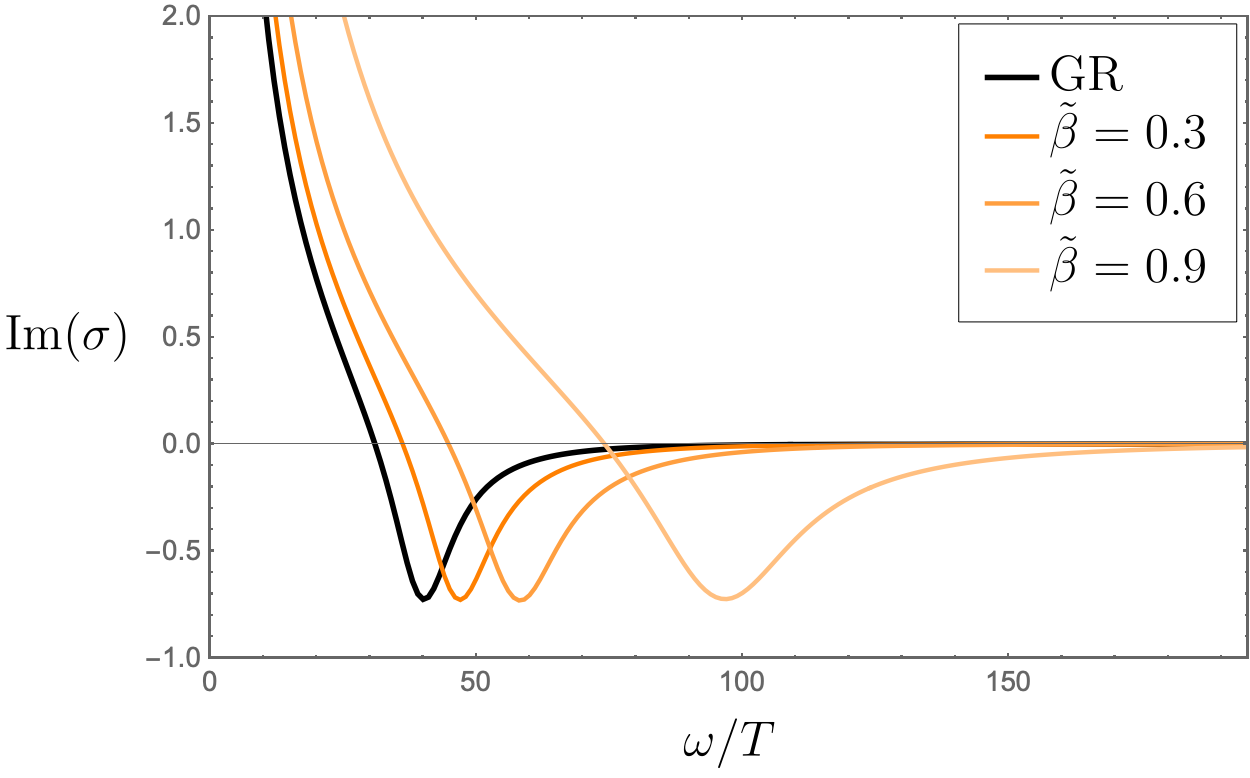}
		\caption*{$\langle \mathcal{O}_+\rangle$ condensate.}
	\end{subfigure}
	\vspace{-0.75cm}
	\caption{Plots of the conductivity as a function of the frequency, for $T/T_c = 0.2$ and different values of $\tilde\beta$. We see that the gap gets larger as $\tilde \beta$ increases. The asymptotic value at large $\omega$ is independent of $\tilde \beta$. }
	\label{fig ECG Conductivity dim1 dim2 beta values T over Tc 02}
\end{figure}

\begin{figure}[h!]
	\centering
	\begin{subfigure}{0.44\linewidth}
		\includegraphics[width=\linewidth]{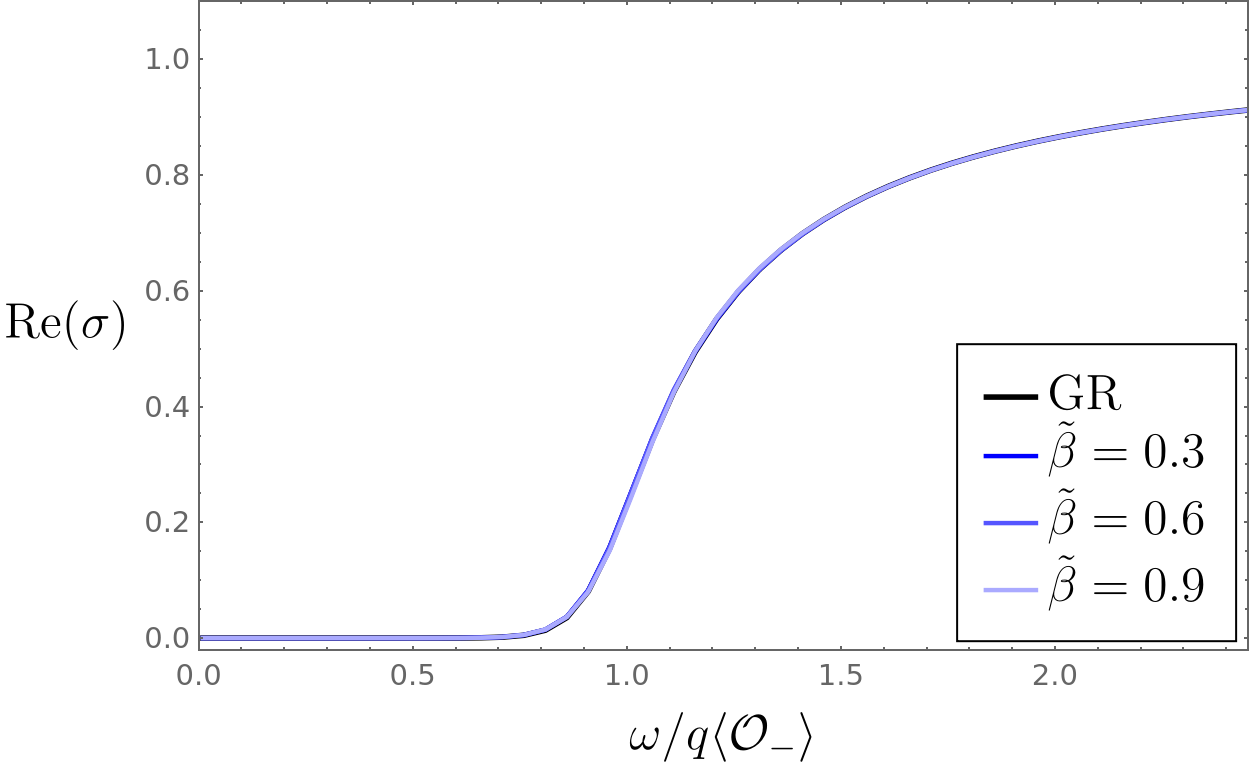}
	\end{subfigure}
	\centering
	\begin{subfigure}{0.44\linewidth}
		\includegraphics[width=\linewidth]{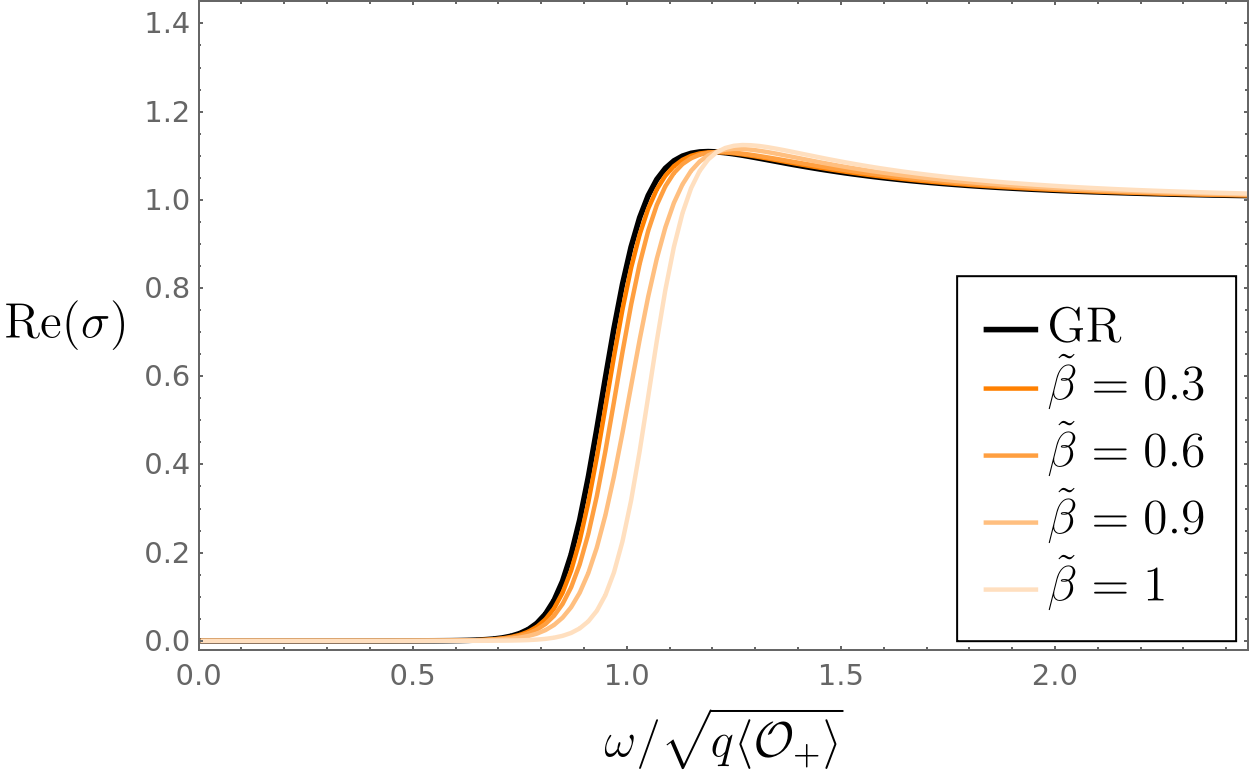}
	\end{subfigure}
	\centering
    \begin{subfigure}{0.44\linewidth}
		\includegraphics[width=\linewidth]{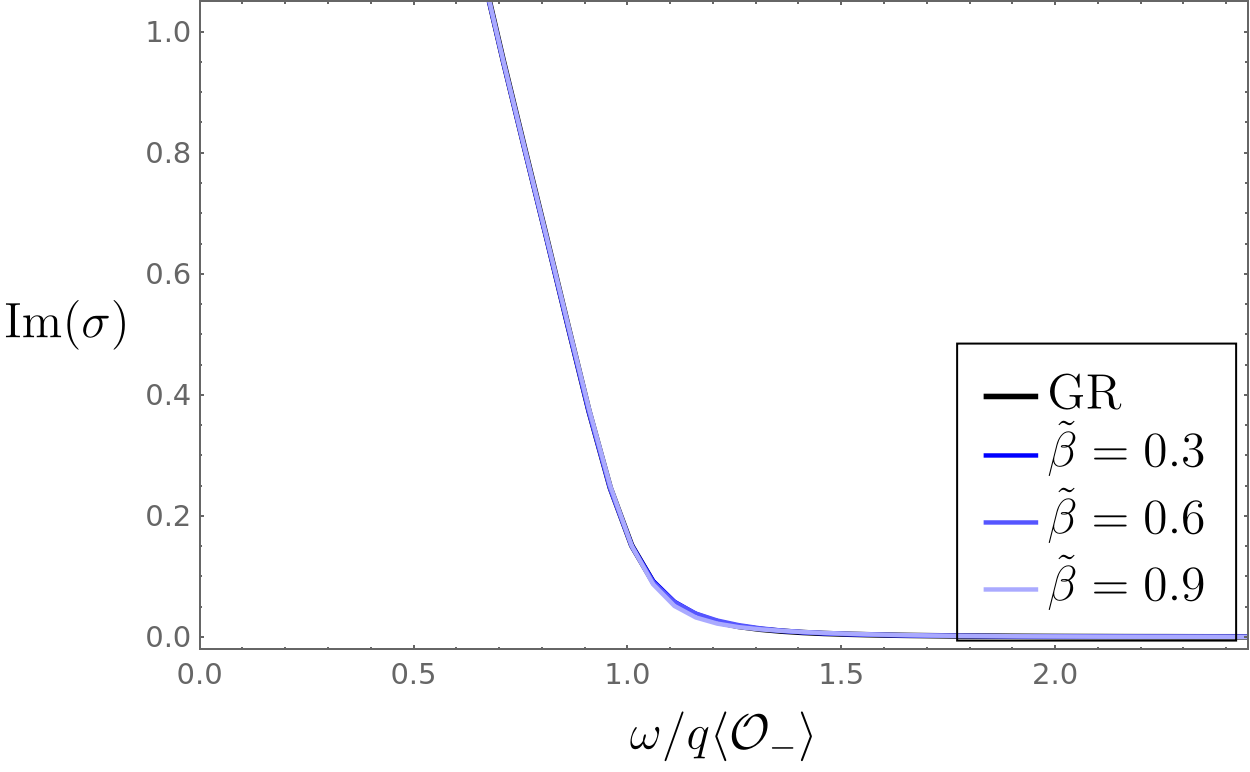}
		\caption*{$\langle \mathcal{O}_-\rangle$ condensate.}
	\end{subfigure}
	\centering
	\begin{subfigure}{0.44\linewidth}
		\includegraphics[width=\linewidth]{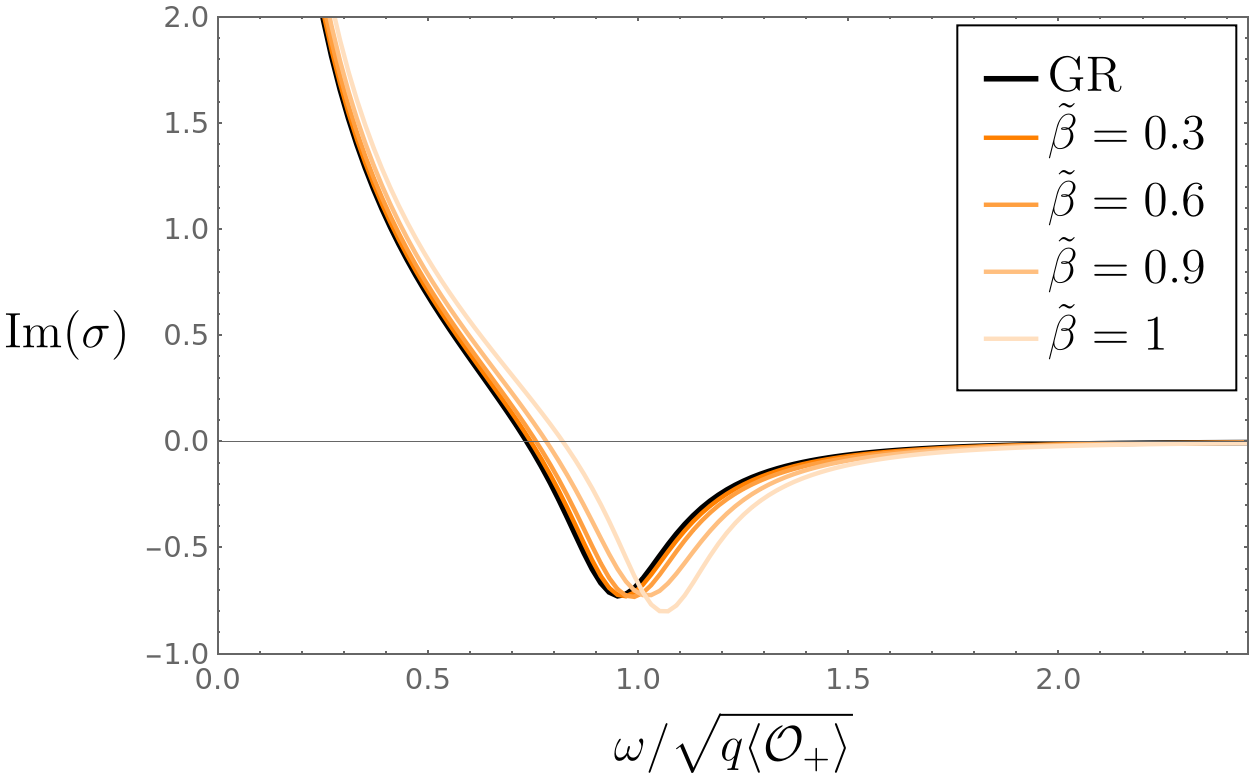}
		\caption*{$\langle \mathcal{O}_+\rangle$ condensate.}
	\end{subfigure}
	\vspace{-.1cm}
	\caption{Plotting the conductivity with a re-scaled $\omega$ axis, we see that the gap dependence on $\tilde \beta$ is mostly due to that of the condensate, the residual part being stronger in the $\langle{\cal O}_-\rangle$ case.}
	\label{fig ECG Conductivity dim1 dim2 beta values T over Tc 02 v2}
\end{figure}

Plots of the resulting conductivities as functions of the frequency are shown in Fig.\ \ref{fig ECG Conductivity dim1 dim2 beta values T over Tc 02}. We observe that they are qualitatively similar to those of the Einstein gravity case $\tilde\beta=0$, the gap getting larger as the cubic coupling $\tilde\beta$ grows. The dependence of the gap on $\tilde\beta$ mimics almost completely that of the condensate, as it can be confirmed when re-scaling the frequency with the appropirate power of condensate value, see Fig.\ \ref{fig ECG Conductivity dim1 dim2 beta values T over Tc 02 v2}. The residual dependence can be atributed to $f_\infty$ in \eqref{eq ECG conductivity formula}, and it is stronger on the $\langle {\cal O}_+\rangle$ case. This could have been expected on the grounds of the larger conformal dimension of the operator, as in eq. \eqref{eq ECG EOM Ax original} the rescaling of $\omega$ makes it appear linearly as $\langle{\cal O}_+\rangle/F^2\sim \langle{\cal O}_+\rangle/f_\infty^2$, as compared to quadratically $\langle{\cal O}_-\rangle^2/f_\infty^2$ in the complementary case. In both cases, the asymptotic value for large frequencies is independent of the cubic coupling.  

\newpage

\vspace{-0.3cm}

\subsection{Backreaction}
\label{sec Backreaction}

Now we turn into the fully backreacting case. To that end, we need to impose the correct boundary conditions for the fields at the horizon and at the boundary. Since the equations of motion \eqref{eq ECG EOMs original fields} have a total of ten derivatives, a naive counting would result in ten constants being required to fully determine the solution. However, several constraints can be found which reduce such number. 

In the standard GR case, we would expect at most a second order equation for each of the four functions on our Ansatz, resulting in eight arbitrary constants. However, coordinate invariance reduces in two the total order of the metric equations. Two of the remaining constants are fixed by demanding $n_\infty^2 {f_\infty}=r_h=1$, and another one by setting $\phi=0$ at the horizon to avoid singularities, as in \eqref{bcphi}. One further reduction is obtained by evaluating at the horizon the equation of motion for $\psi$, which fixes $\psi'(r_h)$ in terms of $\psi_h$. The remaining two constants can be identified with the chemical potential $\mu$ in \eqref{eq:asympPhi} and the value of the condensate $\langle{\cal O}_+\rangle\propto \psi_+$ (aternatively $\langle{\cal O}_-\rangle\propto \psi_-$). When the value of the boundary source ${\cal J}_-\propto \psi_-$ (aternatively ${\cal J}_+\propto \psi_+$) in \eqref{eq:asympPsi} is set to zero, a functional relation is stablished between $\langle{\cal O}_\pm\rangle$ and $\mu$ (or more precisely, the dimensionless combination $\mu/T$).

In the Einsteinian cubic model, the gravitational sixth order system becomes third order after reordering, as it can be checked via a power series expansion at the horizon. This results in a total of seven derivatives. The aformentioned conditions $n_\infty^2 {f_\infty}=r_h=1$ and $\phi(r_h)=0$ reduce in three the number of integration constants. As before, the equation for $\psi$ fixes $\psi'(r_h)$ in terms of $\psi_h$, reducing the total number of arbitrary contants to three. Two of them can be identified with the chemical potential $\mu$ in \eqref{eq:asympPhi} and the value of the condensate $\langle{\cal O}_+\rangle\propto \psi_+$ (aternatively $\langle{\cal O}_-\rangle\propto \psi_-$). After setting to zero the value of the boundary source ${\cal J}_-\propto \psi_-$ (aternatively ${\cal J}_+\propto \psi_+$) and demanding that the solution for $f$ is regular at infinity, we are led to a relation between $\langle{\cal O}_\pm\rangle$ and $\mu/T$.

~ 

\begin{figure}[H]
	\centering
	\begin{subfigure}{0.49\linewidth}
		\includegraphics[scale=0.43]{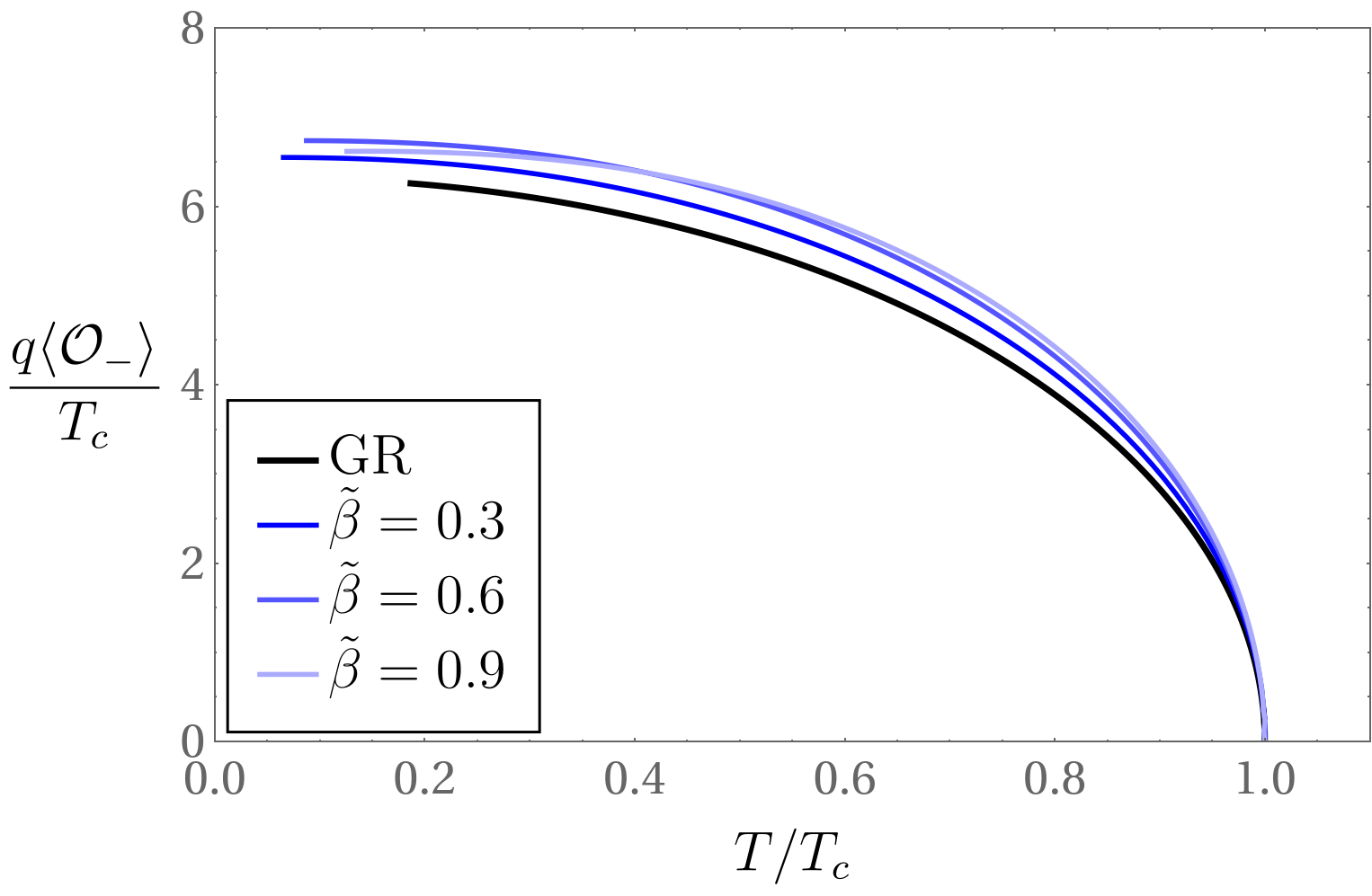}
		\caption*{$\langle \mathcal{O}_- \rangle$ condensate.}
	\end{subfigure}
	\centering
	\begin{subfigure}{0.49\linewidth}
		\includegraphics[scale=0.43]{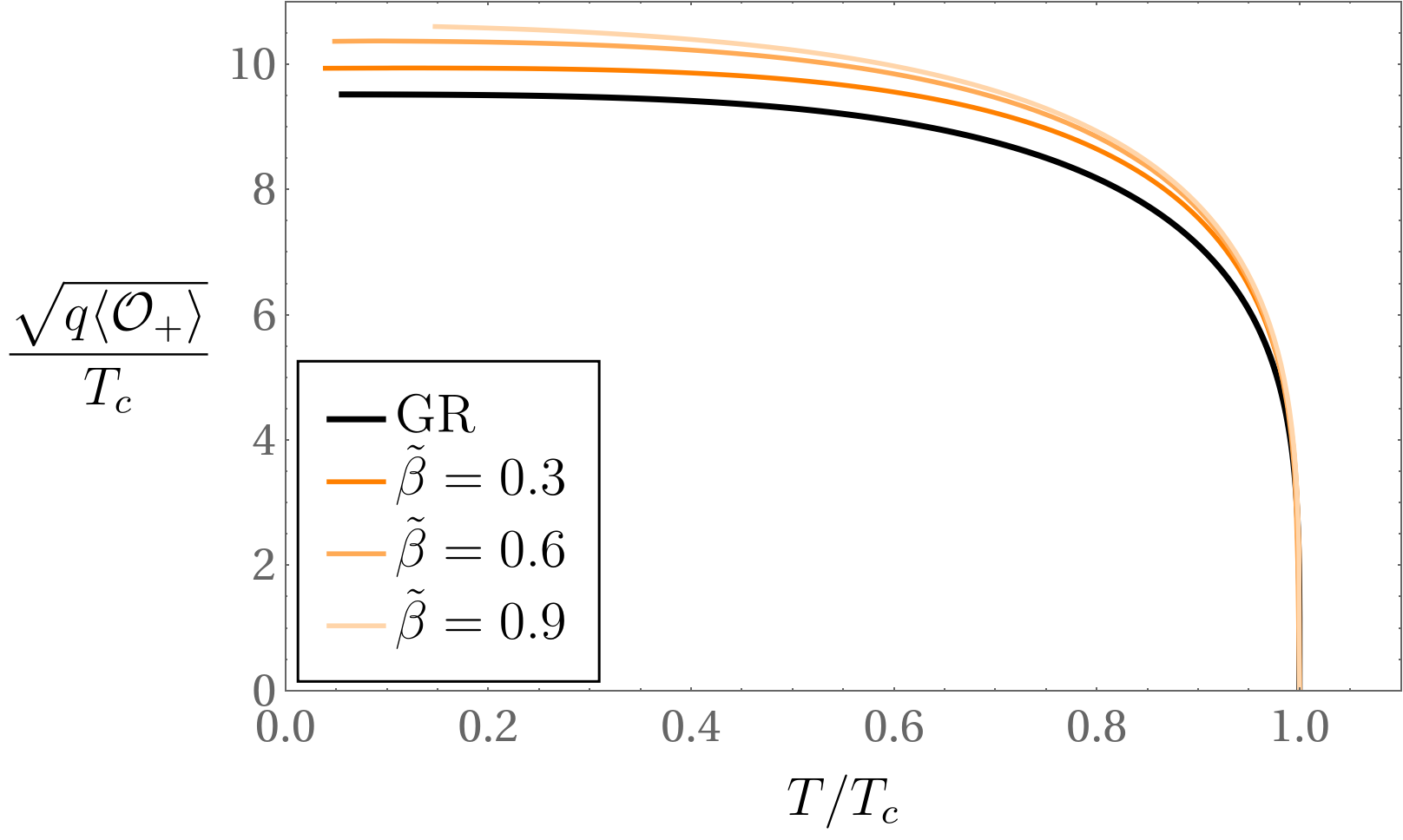}
		\caption*{$\langle \mathcal{O}_+ \rangle$ condensate.}
	\end{subfigure}
	\caption{	\label{fig:backreacted}
	Condensation of the dimension 1 and 2 operators with respect to the temperature in the backreacted case ($q = 3$), for different values of $\tilde\beta$.}
	\label{fig ECG Condensation of the operators with respect to the temperature with q 3. Different beta}
\end{figure}

\begin{figure}[H]
	\centering
	\begin{subfigure}{0.49\linewidth}
		\includegraphics[scale=0.43]{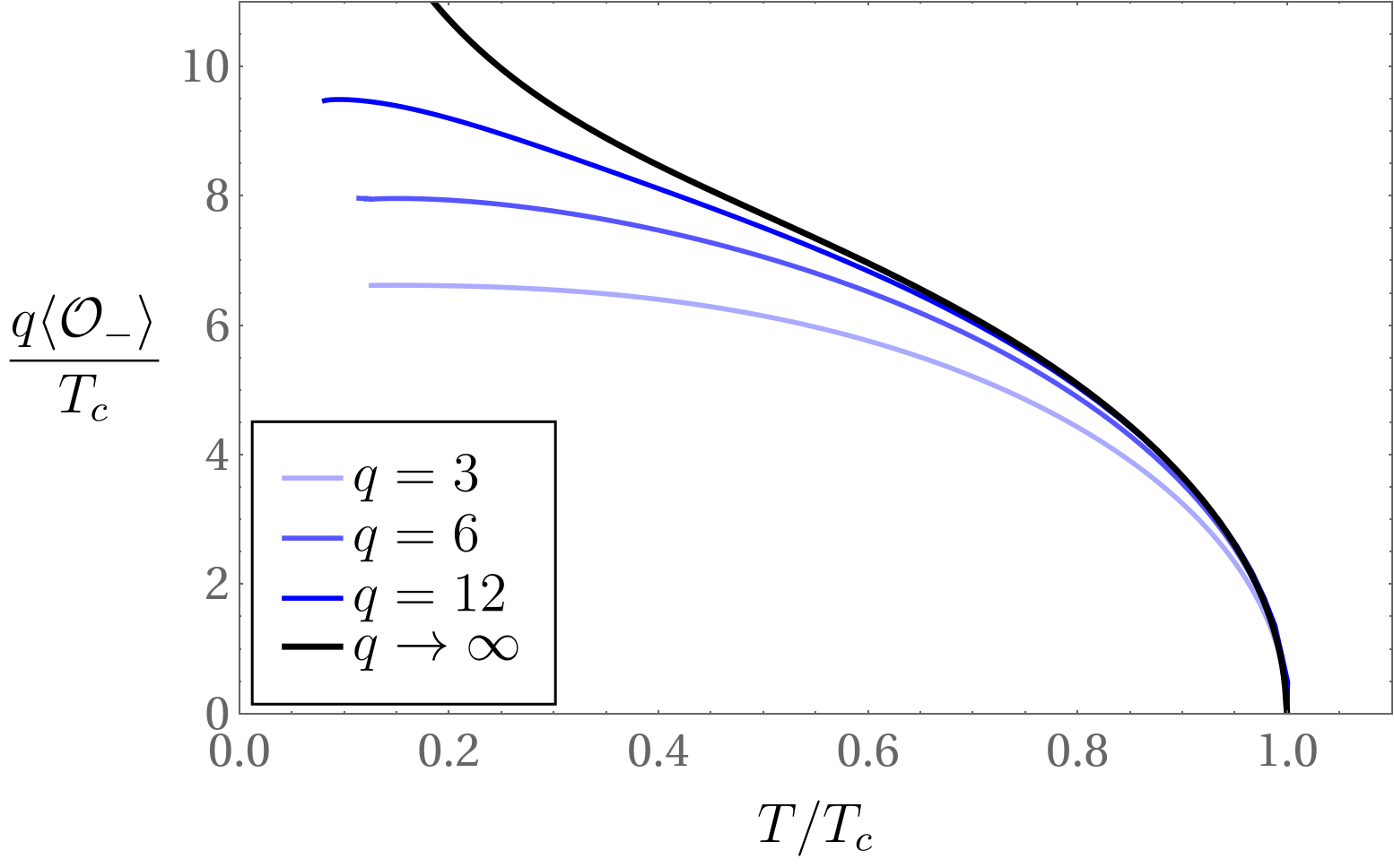}
		\caption*{$\langle \mathcal{O}_- \rangle$ condensate.}
	\end{subfigure}
	\centering
	\begin{subfigure}{0.49\linewidth}
		\includegraphics[scale=0.43]{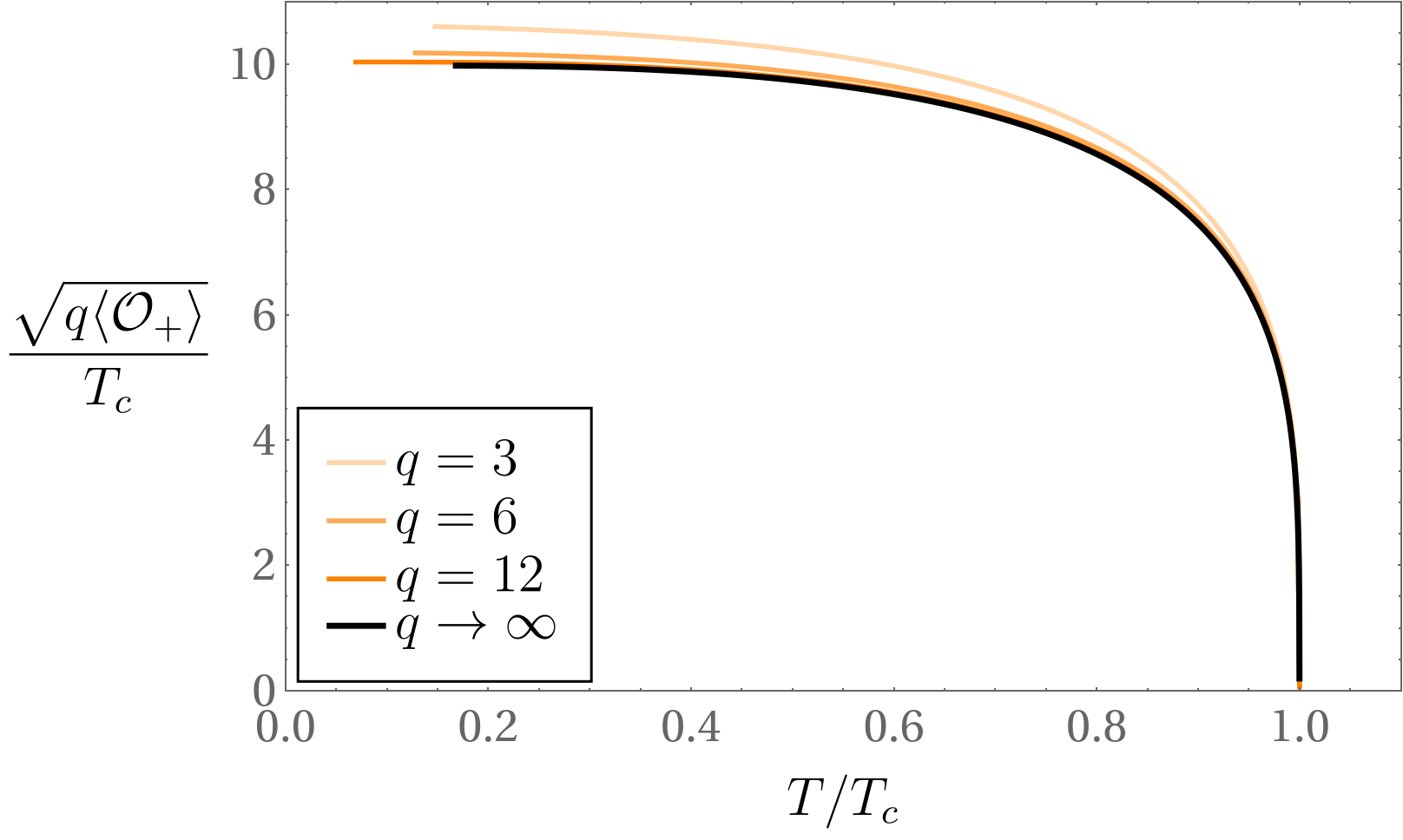}
		\caption*{$\langle \mathcal{O}_+ \rangle$ condensate.}
	\end{subfigure}
	\caption{	\label{fig:backreacted2}
	Condensation of the dimension 1 and 2 operators with respect to the temperature with different strengths of the backreaction $q$, for $\tilde\beta = 0.9$.}
	\label{fig ECG Condensation of the operators with respect to the temperature with beta 0.9. Different q}
\end{figure}

In the plots, we see the condensate $\langle{\cal O}_\pm\rangle$ profiles as a function of temperature for the backreacted case, for different values of $\tilde{\beta}$ (Fig.\ \ref{fig:backreacted}) and of the backreaction $q$ (Fig.\ \ref{fig:backreacted2}). In Fig.\ \ref{fig:backreacted} we see again that larger values of the cubic coupling $\tilde \beta$ lead to larger condensates.  In Fig.\ \ref{fig:backreacted2} we verify that, similarly to the standard Einstein case, the $\langle{\cal O}_+\rangle$  condensate gets larger due to backreaction, while the $\langle{\cal O}_-\rangle$ gets smaller. As advanced, the $\langle{\cal O}_-\rangle$ condensate is now bounded at low temperatures.

\begin{figure}[H]
	\centering
	\begin{subfigure}{0.49\linewidth}
		\includegraphics[width=\linewidth]{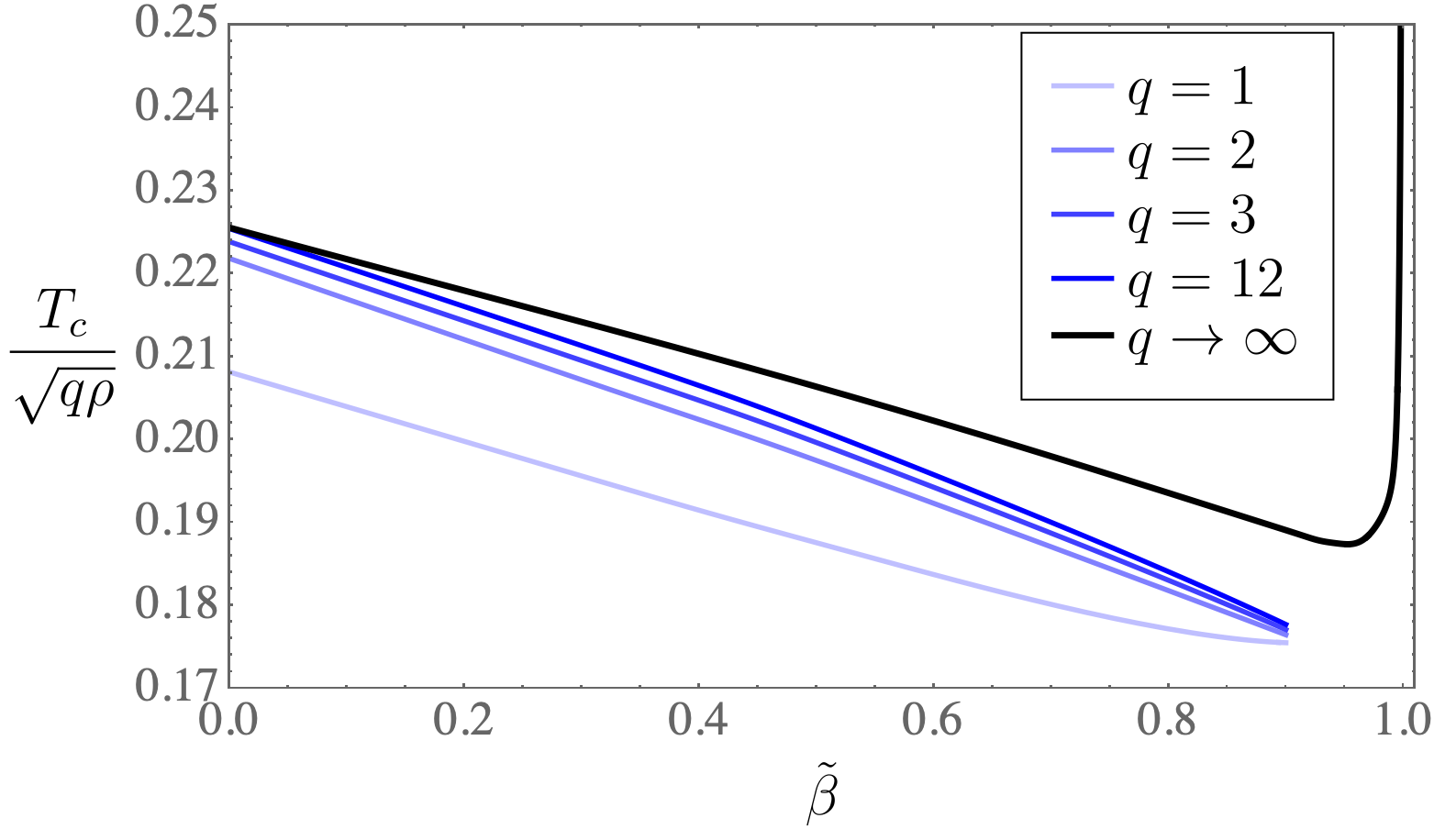}
		\caption*{$\langle \mathcal{O}_1 \rangle$ condensate.}
	\end{subfigure}
	\centering
	\begin{subfigure}{0.49\linewidth}
		\includegraphics[width=\linewidth]{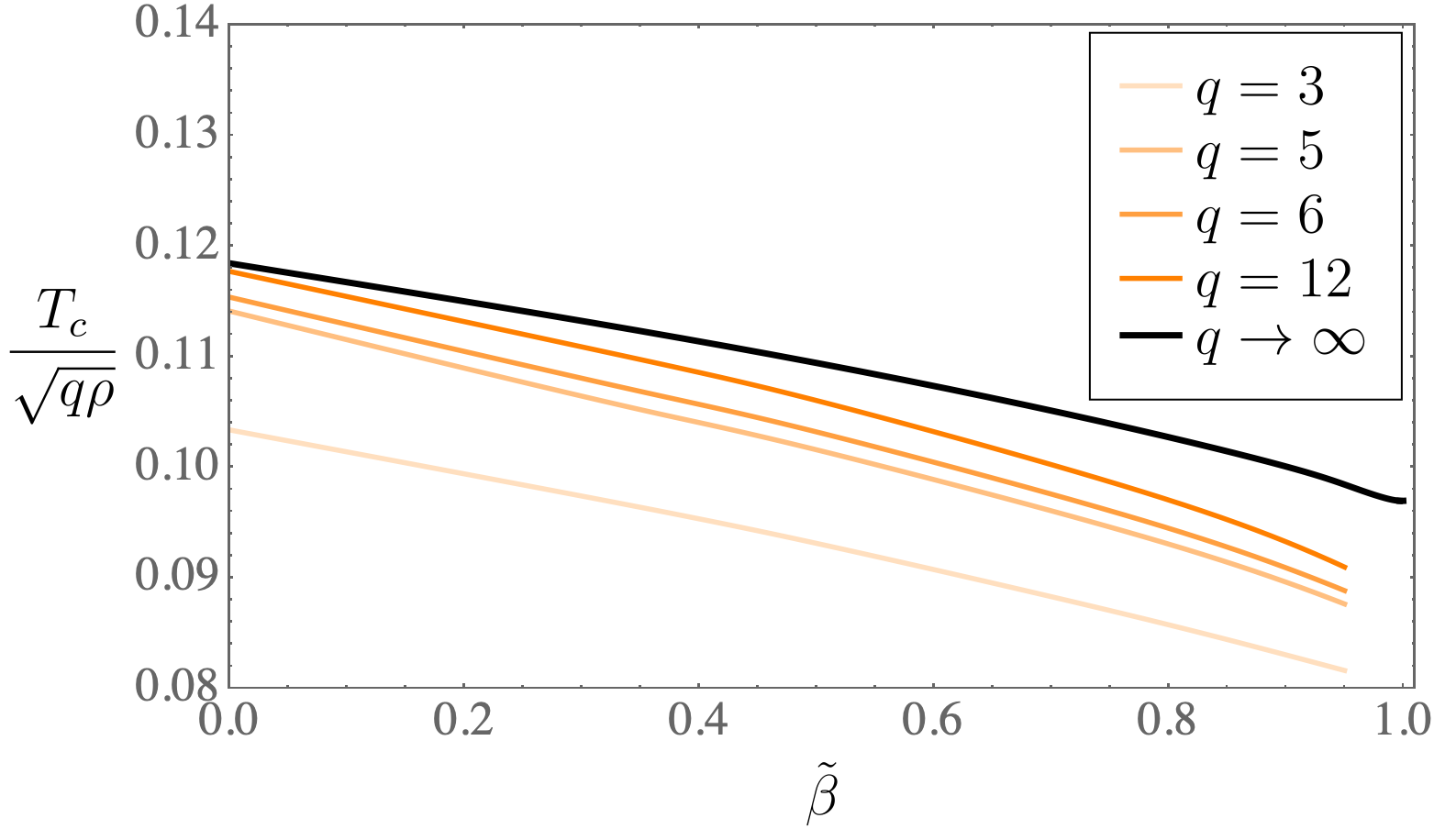}
		\caption*{$\langle \mathcal{O}_2 \rangle$ condensate.}
	\end{subfigure}
	\caption{Critical temperature for the superconducting transition as a function of the cubic coupling $\tilde\beta$, for different values of the backreaction $q$.}
	\label{fig ECG critical temperature with respect to beta. Different q}
\end{figure}

In Fig.\ \ref{fig ECG critical temperature with respect to beta. Different q} we see the behavior of the critical temperature for the superconducting transition, again decreasing monotonically as a function of the cubic coupling, similarly to higher curvature superconductors in higher dimensions \cite{Gregory, Kuang}. As in the standard Einstein case, the critical temperature decreases as backreaction becomes more important.

~ 

\newpage
\section{Discussion and outlook}
\label{sec Discussion}
We investigated the holographic superconductor in $2+1$ dimensions in the context of Einsteinian cubic gravity, with the aim of exploring the effects of finite $N$  and finite 't Hooft coupling corrections on the $2+1$ dimensional boundary physics. In particular, we studied the dependence the critical temperature of the phase transtion and on the strength of the resulting condensate, as well as the electric conductivity.

We found that, as previously reported in the higher dimensional Gauss-Bonnet case \cite{Gregory,Pan, Kuang,Barclay}, the presence of higher curvature corrections to the bulk action makes the condensation of the scalar more difficult, lowering the critical temperature. When the critical temperature is reached and the condensation occurs, we find that the resulting condensate is larger as the cubic coupling $\tilde\beta$ grows, both in the probe limit and in the fully backreacting case. As in the standard case, there is a gap in the electric conductivity as a function of the frequency, whose size is compatible with the value of the condensate. A mild residual dependence of the conductivity gap on $\tilde \beta$ can be attibuted to the asymptotic value of the black hole radial function.

The present result could be straigthfordwardly expanded in the direction of including higher order terms in the bulk, along the lines of the so-called generalized quasitopological theories. 

An interesting point about the present results is the following. The decrease of the critical temperature as the higher curvature coupling is increased could be naively attributed to a gradual restoration of the validity of the Coleman-Mermin-Wagner theorem, as $1/N$ corrections are included. Indeed, in contrast with the previously known higher curvature superconductors \cite{Gregory, Pan, Kuang, Barclay, color}, in the present Einsteinian cubic gravity context the boundary field theory is $2+1$ dimensional, satifying one of the main theorem's hypotheses.
However, higher curvature corrections to the gravitational action arise from bulk loop diagrams which are finite in the IR, while Coleman-Mermin-Wagner result is due to an IR divergence on the Goldstone boson correlator \cite{Beekman}. Thus, these are different effects which are not related, the decrease of the critical temperature must then be attributed to different physics. Since it seems to be a generic behavior, it deserves further investigation.

\section*{Acknowledgements}

We would like to thank Sean Hartnoll for helpful correspondence.
The work of JDE and ARS was supported by AEI-Spain (under project PID2020-114157GB-I00 and Unidad de Excelencia Mar\'\i a de Maeztu MDM-2016-0692), by Xunta de Galicia-Conseller\'\i a de Educaci\'on (Centro singular de investigaci\'on de Galicia accreditation 2019-2022, and project ED431C-2021/14), and by the European Union FEDER.
That of NEG was suported by CONICET grants PIP-2017-1109 and PUE084 ``B\'usqueda de nueva f\'\i sica'', and UNLP grant PID-X791, and he is grateful to Universidad de Concepci\' on and Centro de Estudios Cient\' \i ficos for hospitality and support during the late part of this work.
ARS is supported by the Spanish MECD fellowship FPU18/03719, and is pleased to thank KU Leuven, where part of this work was done, for their warm hospitality.

\newpage
\appendix

\section{Numerical relaxation method for differential equations}
\label{appendix Numerical relaxation method for differential equations}

We will now explain the basics of the relaxation method \cite{Tomas Andrade Relaxation method}, used throughout the computations described in the main text. The basic goal of this method is to convert an arbitrarily complicated system of differential equations into a matrix equation, which can be solved using linear algebra methods.

Suppose a linear differential equation of the form
\begin{equation}
	L[y(x)] = J(x), \quad \text{with} \quad x \in [a, \, b] \,.
\end{equation}
The interval of values of the independent variable $x$ can be discretized in $N$ points $x_i$, with $i = 1, \dots , N+1$. We can therefore construct a vector $\vec{y}$ such that $y_i \equiv y(x_i)$, and the same thing for the source term $J(x)$. In this discrete form, derivative operators of any order $n$,  $\partial_x^n$, can be written as $(N+1)\times(N+1)$ matrices\footnote{These operators are constructed automatically by a function in Mathematica, which yields a matrix that computes the order $n$ derivative at each point using an arbitrary number of nearby points.} $D_n$. It is also possible to replace an order $n$ derivative operator by the product of $n$ first-derivative matrices,
\begin{equation}
	\partial_x^n \, \longrightarrow \, D_n = (D_1)^n \,.
\end{equation}
If the operator $L[y(x)]$ contains a term of the form $f(x) y(x)$, this $f(x)$ should be replaced by a discrete version $F$, given by the matrix
\begin{equation}
	F = \text{diag} \left( f(x_1), \dots , f(x_{N+1}) \right) \,.
\end{equation}
If this appears as $f(x) \partial_x^n y(x)$, it is replaced by the product $F D_n$. After doing all the substitutions, the total operator $L$ is transformed into a matrix $M$ by adding each of the individual terms of the differential equation together. The final system to solve is
\begin{equation}
	M \cdot \vec{y} = \vec{J} \,.
\end{equation}
But we also need to apply the boundary conditions. These are enforced by modifying the components of $M$ and $\vec{J}$.
\begin{enumerate}
	\item If we have a condition of the form $y(x_j) = A$, set
	\begin{equation}
		M_{ij} = 0 \quad \text{if} \quad i \neq j \,, \qquad M_{jj} = 1 \,, \qquad J_j = A \,.
	\end{equation}
	\item If the condition is of the form $\partial_x^n y(x_j) = B$, set
	\begin{equation}
		M_{j} = (D_n)_{j} \,, \qquad J_j = B \,.
	\end{equation}
\end{enumerate}
The described implementation works for linear differential equations, as it relies on the use of linear operators. The generalization to non-linear ODEs is slightly more involved, but also possible.

Consider a non-linear differential equation
\begin{equation}
	E[y(x)] = J(x) \,.
	\label{eq appendix nonlinear ODE}
\end{equation}
The solution can be found iteratively starting from an initial seed $y_0$. We expand the solution as $y(x) = y_0(x) + \delta y(x)$, where $\delta y(x)$ is the change on the solution after one iteration. After replacing this in the equation (\ref{eq appendix nonlinear ODE}) and expanding to linear order in $\delta y(x)$ it becomes
\begin{equation}
	E[y_0(x)] + \delta E[\delta y(x)] = J(x) \,.
\end{equation}
Therefore, $\delta E[\delta y(x)]$ is a linear differential operator acting on $\delta y(x)$, and $E[y_0(x)]$ should be considered part of the inhomogeneous term. $\delta y(x)$ can now be computed using the method described above, and at the end it is necessary to update the total solution $y(x)$, which becomes the seed for the following iteration. The process is repeated until the solution converges.

The method can be straightforwardly generalized for coupled differential equations. For this, it is enough to concatenate the vectors formed by discretizing every function $y^{(a)}(x)$ in one single vector,
\begin{equation}
	\vec{y} = \left( \vec{y}^{(1)}, \vec{y}^{(2)}, \dots, \vec{y}^{(n)} \right) \, .
\end{equation}
The matrix $M$ also becomes more complex, since now each differential equation can depend on all the functions involved. It can be constructed by blocks, with each block corresponding to one of the equations in the system, following the same procedure as before. Once this is done, the rest of the computation carries on as usual.

The method can be generalized for solving also partial differential equations. However, since this is not necessary for this work we do not explain it here, and we refer to \cite{Tomas Andrade Relaxation method}.
 
\section{Deep IR geometry and effective mass }
\label{appendix AdS2}
In this section, we want to explore whether the condensation becoming more difficult as $\tilde{\beta}$ increases can be explained by studying the system at very low temperatures. For that, let us take the scalar field $\psi$ to have a mass $m$ and a vanishing value, since we want to study the onset of the condensation.

Let us assume the background to be an AdS Reissner-Nordström black hole, with a metric of the form \eqref{eq 2 Ansatz planar black hole} and a background electrostatic potential $\phi(r)$, as in the main text. The difference here is that this Maxwell field will influence the form of the metric. Since $\psi = 0$, the equation of motion of this potential \eqref{eq:gauge} reduces to
\begin{equation}
	\frac{d}{dr} \left( \frac{r^2 \phi' (r)}{n(r)} \right) = 0 ~,
\end{equation}
which is solved by
\begin{equation}
	\phi' (r) = Q \frac{L^2}{r^2} n(r) ~,
\end{equation}
where $Q$ is a constant, proportional to the electric charge of the black hole. This ansatz also admits a single-function solution, so we can set $n(r) = n_\infty$, which we will fix later. Then, the equation of motion for the function $F(r)$ reads
\begin{equation}
\frac{r^3}{L^2} - r F + \frac{L^4 \kappa^2 Q^2}{2 r} + \frac{L^4}{27} \tilde{\beta} \left( 3 F F' F'' - F'^3 + \frac{6 F^2}{r^2} \left( F' - r F'' \right) \right) = M ~.
	\label{eq:app EOM F(r)}
\end{equation}
Studying the asymptotic limit $r \rightarrow \infty$, we see that the function $F(r)$ behaves as before, $F(r) \simeq r^2 f_\infty / L^2$, with $f_\infty$ given by \eqref{eq 2 f_infinity analytic expression}, so we also fix $n_\infty = 1 / \sqrt{f_\infty}$. Let us now expand $F(r)$ near the horizon as
\begin{equation}
	F(r) =\frac{4 \pi T}{n_\infty} (r - r_h) + \frac{1}{\ell^2} ( r - r_h)^2 +{\cal O}((r-r_h)^3)~,
	\label{eq:app Expansion F(r) near r_h}
\end{equation}
where $\ell$ is a constant with units of length. Plugging this in \eqref{eq:app EOM F(r)} we find the relations
\begin{equation}
	T = \frac{6 r_h^4 - L^6 \kappa^2 Q^2}{8 \pi L^2 r_h^3 \sqrt{f_\infty}} ~, \qquad M = \frac{r_h^3}{L^2} + \frac{L^4 \kappa^2 Q^2}{2 r_h} - \frac{\tilde{\beta}}{L^2} \left( \frac{6 r_h^4 - L^6 \kappa^2 Q^2}{6 r_h^3} \right)^3 ~.
\end{equation}
Since we are interested on the behavior of the system at very low temperatures, we will take the extremal limit of this, $T = 0$, which corresponds to
\begin{equation}
	Q^2 = \frac{6 r_h^4}{L^6 \kappa^2} ~, \qquad r_h = \left( \frac{M L^2}{4} \right)^{1/3} ~.
	\label{eq:app Constants extremal limit}
\end{equation}
In this case, it is easy to obtain the value of $\ell$ in the expansion \eqref{eq:app Expansion F(r) near r_h}, by replacing it again in the equation \eqref{eq:app EOM F(r)} and using the values of $Q$ and $r_h$ above, finding
\begin{equation}
	\ell^2 = \frac{L^2}{6} ~,
	\label{eq:app Value of ell in the extremal limit}
\end{equation}
which is independent of $\tilde{\beta}$. 

Let us now introduce the coordinates
\begin{equation}
	r - r_h = \epsilon \frac{\ell^2}{L^2} \rho ~, \qquad t = \frac{L^2}{\ell^2} \frac{\tau}{\epsilon n_\infty} ~,
	\label{eq:app Change of coordinates near-horizon extremal RN black hole}
\end{equation}
chosen in such a way that the near-horizon metric becomes
\begin{equation}
	ds^2 = - \frac{\rho^2}{\ell^2} d\tau + \frac{\ell^2}{\rho^2} d\rho^2 + r_h^2 (dx^2 + dy^2) ~,
	\label{eq:app Near-horizon metric extremal RN black hole}
\end{equation}
which corresponds to the product of AdS$_2$ with the real plane. The electrostatic potential $\phi$ is also modified by this choice of coordinates, and it reads\footnote{
This can be seen by writing $A = \phi dt = \phi_\tau d\tau$, where the coordinates $t$ and $\tau$ are related as in \eqref{eq:app Change of coordinates near-horizon extremal RN black hole}, and we expand $\phi(r)$ near $r = r_h$ as $\phi(r) \simeq \phi'(r_h) (r - r_h)$.
}
\begin{equation}
	\phi_\tau \simeq Q \frac{L^2}{r_h^2} \rho ~.
\end{equation}
The equation of motion of a scalar field $\psi(\rho)$ of mass $m$ in this background is
\begin{equation}
	\psi'' + \frac{2}{\rho} \psi' + \frac{\ell^2}{\rho^2} \left( \frac{\phi_\tau^2 \ell^2}{\rho^2} - m^2 \right) \psi = 0 ~,
\end{equation}
so the scalar field has an effective mass (squared) given by the second parenthesis, which in the extremal limit \eqref{eq:app Constants extremal limit} reads
\begin{equation}
	m_{\rm eff}^2 = m^2 -  \frac{6 \ell^2}{L^2 \kappa^2} ~.
\end{equation}
This is independent of $\tilde{\beta}$, since $\ell^2$ is given by \eqref{eq:app Value of ell in the extremal limit}. In order for the scalar field to become unstable and condense at these low temperatures, it would need to violate the BF bound, which in our AdS$_2$ spacetime reads
\begin{equation}
	m_{\rm eff}^2 \geq - \frac{1}{4 \ell^2} ~.
\end{equation}
Equivalently, for the original mass $m$ this bound implies
\begin{equation}
	m^2 \geq   \frac{6 \ell^2}{L^2 \kappa^2} - \frac{1}{4 \ell^2} ~.
	\label{eq:app Effective BF bound for the extremal RN black hole}
\end{equation}
This effective BF bound is independent of the coupling of the cubic terms $\tilde{\beta}$. However, recall that in the main text we had fixed the value of $L^2 m^2 / f_\infty=-2$ in order for the scalar field to have the desired behavior near the boundary. This implies that the bound above reads, after plugging in also the value of $\ell^2$ in \eqref{eq:app Value of ell in the extremal limit} and setting $L = 1$ and $2 \kappa^2 = 1$ as in the main text,
\begin{equation}
	-
{f_\infty}
\geq \frac{1}{4} ~.
	\label{eq:app Effective BF bound for the extremal RN black hole 2}
\end{equation}
Since $f_\infty$ is a positive number that grows with $\tilde \beta$ (see Fig.\ \ref{fig 2 Numerical solution black hole several beta}), 
this inequality is always violated, leading to the superconducting instability. 

\end{document}